\documentclass[journal=jacsat,manuscript=article, natbib=true]{achemso}
\setcitestyle{numbers,square}
\usepackage[version=3]{mhchem}
\usepackage{adjustbox}
\usepackage{threeparttable}
\usepackage{booktabs}
\usepackage[font=bf]{caption}
\usepackage{graphicx}
\usepackage{subcaption}  % modern replacement for subfigure
\usepackage{rotating}

\SectionNumbersOn

\usepackage[colorlinks=true,
            linkcolor=blue,
            citecolor=magenta,
            urlcolor=blue]{hyperref}
\author{Sudipta Chakraborty}
\author{Kamal Majee}
\author{Achintya Kumar Dutta}
\email{achintya@chem.iitb.ac.in}
\affiliation[Unknown University]
{Department of Chemistry, Indian Institute of Technology Bombay, Mumbai 400076, India}
\alsoaffiliation[Second University]
{Department of Inorganic Chemistry, Faculty of Natural Sciences, Comenius University, Bratislava
Ilkovičova 6, Mlynská dolina 842 15 Bratislava, Slovakia}

\title[An \textsf{achemso} demo]
  {A Low Cost Relativistic Algebraic Diagrammatic Construction Method Based on Cholesky Decomposition and Frozen Natural Spinors for Electronic Ionization, Attachment and Excitation Energy Problem}

\abbreviations{IR,NMR,UV}

\begin{document}



\newpage
\begin{abstract}

We present an efficient relativistic implementation of algebraic diagrammatic construction (ADC) theory up to third order for the treatment of electronic ionization potentials (IP), electron affinities (EA), and excitation energies (EE) in heavy-element systems using an exact two-component atomic mean-field (X2CAMF) Hamiltonian. The approach combines Cholesky decomposition (CD) of two-electron integrals with frozen natural spinors (FNS) to significantly reduce the computational cost without compromising accuracy. To improve the description of excited states, we have implemented a state-specific frozen natural spinor (SS-FNS) framework and applied it to both electron affinity and excitation energy calculations. In addition to the standard relativistic ADC(3) method, we investigate a semi-empirically scaled variant in which the third-order contribution to the ADC secular matrix is multiplied by a scaling factor (x), denoted as FNS/SS-FNS-[ADC(2)+(x)(3)]. This [ADC(2)+(x)(3)] approach shows systematic improvements over conventional ADC(3) in a variety of cases. Substantial computational savings are achieved through the use of FNS and SS-FNS schemes when compared to canonical calculations, resulting in significant speedups for ionization, attachment, and excitation energy computations. The current implementation accurately reproduces the canonical four-component ADC(3) results while significantly reducing computational cost. The efficiency and robustness of the method are demonstrated through applications to medium and large-sized molecular systems, including systems with 70 atoms and over 2600 basis functions.

\end{abstract}

\newpage
\section{Introduction}
The accurate and efficient simulation of electronic excitation, ionization, and electron attachment processes remains a cornerstone for understanding a wide range of chemical and physical phenomena, such as photochemical reactions, electronic spectroscopy, charge-transfer processes, and photoionization.\cite{gonzalez2020quantum} Numerous methods have been developed for evaluating ionized, electron-attached, and excited-state energies, which can be broadly classified into two categories. The first category, known as ``$\Delta-$ based methods," involves performing two separate total energy calculations, one for the ground state and another for the target state, and subsequently taking their difference to obtain the desired energy. The other approach is the “direct energy-difference method,” which requires only a single calculation from which excitation, ionization, or electron attachment energies are obtained directly as eigenvalues of an effective secular Hamiltonian.  The former approach often suffers from issues like symmetry breaking and variational collapse of the wave function, which can compromise the reliability of the computed energy differences. In addition to these challenges, it also lacks access to important transition properties, including spectroscopic amplitudes or intensities. In contrast, the latter approach is generally free from these problems and is therefore more commonly used in modern electronic structure calculations. \\
The equation-of-motion coupled cluster (EOM-CC)\cite{krylov2008equation,nooijen1992coupled,stanton1993equation,rowe1968equations} method has become one of the most widely used direct energy-difference wave function methods for computing excitation energies (EE), electron affinities (EA), and ionization potentials (IP).  The linear-response coupled cluster (LR-CC)\cite{koch1990excitation,sekino1984linear} method yields identical EE, although it is formulated from a different theoretical perspective than the EOM-CC approach.
However, in both cases, the non-Hermitian nature of the effective Hamiltonian within the biorthogonal coupled-cluster framework renders the eigenvalue problem susceptible to complex solutions. \cite{thomas2021complex} Moreover, the computation of transition properties within standard coupled-cluster (CC) theory is challenging, as it requires the determination of both left and right eigenvectors. The unitary coupled cluster (UCC) approach offers a Hermitian formulation of CC methods and helps address these challenges.\cite{liu2018unitary,liu2021unitary} However, the non-terminating Baker-Campbell-Hausdorff (BCH) expansion of the similarity-transformed Hamiltonian poses significant challenges for the practical implementation of the UCC method. To address this issue, commutator-based truncation schemes within UCC theory have garnered considerable attention, particularly due to their relevance in quantum computing.\cite{Evangelistaucc_a} Importantly, these schemes have also been shown to be advantageous on classical computers.\cite{majee2024reduced,majee2025relativistic}

In contrast to the iterative nature of CC–based methods, the algebraic diagrammatic construction (ADC)\cite{schirmer2004intermediate,mertins1996algebraic,schirmer1991closed,schirmer1982beyond,von1984computational,dreuw2015algebraic,dempwolff2019efficient,banerjeeEfficientImplementationSinglereference2021} scheme has attracted considerable attention due to its low-cost, non-iterative treatment of the ground-state wave function based on Møller–Plesset perturbation theory (MPn), as well as its Hermitian formulation, which provides distinct advantages for the calculation of excited-state properties. The ADC method was first developed using the propagator formalism.\cite{schirmer1982beyond,trofimov1999consistent}  Propagator-based ADC methods were successfully applied to compute electron-ionization and attachment spectra\cite{fetter2012quantum,cederbaum1975one,cederbaum1977theoretical,schirmer1978two}. In modern implementations of ADC, the intermediate-state representation (ISR) has emerged as the preferred approach, largely due to its transparent formalism and its provision of direct access to the excited-state wavefunction.\cite{schirmer1991closed} An extension of the fourth-order ADC (ADC(4)) scheme within the ISR formalism for EE, IP, and EA has been proposed by Dreuw and co-workers.\cite{leitner2022fourth, leitner2024fourth} In recent years, ADC methods have been extensively employed for energy and property calculations, particularly for systems containing lighter elements.\cite{leitner2024fourth,schirmer1983new,angonoa1987theoretical, dreuw2015algebraic, trofimov1995efficient, trofimov1997polarization, trofimov1999consistent,schirmer2004intermediate, trofimov2005molecular}
One of the most straightforward ways of introducing relativistic effects in quantum chemical calculations is to employ a four-component (4c) Dirac–Coulomb (DC) Hamiltonian.\cite{dyall2007introduction} Relativistic formulations of ADC at the second-order and extended second-order levels have been introduced by Pernpointner and co-workers to enable the calculation of excitation energies. \cite{pernpointner2014relativistic,pernpointner2018four} In addition, non-Dyson-based second- and third-order ADC schemes have been developed within a relativistic framework to compute valence ionization energies and investigate electron decay processes.\cite{pernpointner2004one,pernpointner2010four,pernpointner2005effect,mandal2025third} Although second-order ADC (ADC(2)) can provide qualitative agreement, it typically lacks quantitative accuracy. To achieve more reliable results, it is necessary to go beyond second-order and include at least third-order contributions in the perturbative expansion. To address this, we have recently developed a four-component (4c) ADC(3) method for calculating IP, EA, EE, and excited state properties of atoms and molecules containing heavy elements.\cite{chakraborty2025relativistic} However, performing 4c-ADC(3) calculations remains a significant challenge due to their extremely high computational cost, which can be up to 32 times higher than that of a non-relativistic method.\cite{dyall2007introduction} As a result, such calculations are typically limited to atoms and small molecules and require moderate-sized basis sets. To mitigate this limitation, numerous approaches have been developed in the literature to lower the computational cost of relativistic electronic structure calculations. One of the most effective strategies is the use of two-component relativistic theories.\cite{dyall2007introduction,hess1986relativistic,van1996relativistic,dyall1997interfacing,nakajima1999new,liu2009exact,saue2011relativistic}  Among the various flavors of two-component methods available, the exact two-component (X2C) theory based on atomic mean-field (AMF) spin–orbit integrals (X2CAMF) has emerged as one of the most attractive approaches.\cite{liu2018atomic,zhang2022atomic,knecht2022exact} It can be further combined with Cholesky decomposition (CD) of the two-electron integrals.\cite{helmich2019relativistic, banerjee2023relativistic, uhlirova2024cholesky}  Sokolov and co-workers\cite{Sokolov} have recently reported a single-reference and multi-reference ADC methods for ionization and electron attachment problems based on a one-electron variant of the X2C method (X2C-1e).  In addition to the simplification offered by the two-component framework, leveraging massively parallel programs can accelerate the floating-point operations of relativistic electron correlation methods.\cite{pototschnig2021implementation}  Alternatively, one can use a more compact set of spinor space in the form of frozen natural spinors (FNS), which allows one to perform effective truncation of the virtual space with systematic and controllable accuracy. The FNS are generally constructed from MP2 calculations, and the efficiency of the MP2-based FNS has been demonstrated in relativistic four-component ground-state coupled cluster (CC)\cite{chamoli2022reduced,chamoli2025frozen} and unitary coupled cluster (UCC)\cite{majee2024reduced} calculations.  This approach has also been extended to core and valence ionization problems\cite{surjuse2022low,chamoli2025reduced} using the equation of motion coupled cluster approach (EOM-CC). Gomes and co-workers\cite{yuan2022assessing} have reported a similar natural spinor-based strategy for the ground-state CC method. To further reduce the computational cost, the FNS method can be combined with the CD-based X2CAMF-CC method.\cite{chamoli2025frozen} Gomes and coworkers\cite{yuan2022assessing} have demonstrated that, in contrast to ground state energy calculations, standard MP2-based natural spinors fail to describe excited states due to their inability to capture the corresponding electronic distribution. It has recently been demonstrated that a state-specific approach can mitigate this problem.\cite{mukhopadhyay2025reduced}  
%In this context, our group has already developed state-specific frozen natural orbitals for ionized-state calculations using ADC in the non-relativistic regime\cite{mukhopadhyay2023state}, as well as for both ionized\cite{manna2024efficient} and excited states\cite{mukhopadhyay2025reduced} within the EOM-CCSD framework in the non-relativistic and relativistic domains, respectively.
In this work, we present the theoretical formulation, computational implementation, and systematic benchmarking of a low-cost relativistic ADC(3) approach for ionized, electron-attached, and electronically excited states, combining the X2CAMF Hamiltonian with CD and the FNS or SS-FNS approximation.

\section{Theory}
\subsection{Relativistic Algebraic Diagrammatic Construction Theory}
The ADC method offers a systematic scheme for the calculation of generalized excitation energies \cite{schirmer1982beyond,dreuw2015algebraic,schirmer2018many,schirmer1991closed,oddershede1978polarization, ODDERSHEDE198433,oddershede1987propagator,schirmer1983new,ortiz2013electron} within the framework of propagator theory. The type of spectroscopic information obtained depends on the propagator under consideration. For instance, employing ADC in connection with the polarization propagator provides access to electronic excitation spectra,\cite{schirmer1982beyond,dreuw2015algebraic,schirmer2018many,trofimov1995efficient,trofimov1997polarization,trofimov1997polarizationII}, whereas  formulations based on the electron propagator enables the calculation of ionization potentials and electron affinities.\cite{schirmer2018many,schirmer1983new,von1984computational,schirmer1989green,schirmer1998non,trofimov2005molecular}

 In the original formulation, the method did not provide access to explicit excited-state wave functions; instead, only excitation energies and transition strengths could be extracted from the poles and residues of the propagator, respectively. This limitation was later resolved through the development of the ISR framework.\cite{mertins1996algebraic,dreuw2015algebraic,schirmer2018many,schirmer1991closed,mertins1996algebraicII} The construction of ISR typically begins with the correlated excited states (CES),
\begin{equation}
|\Psi_{J}^{0} \rangle = \hat{C_{J}}|\Psi_{0} \rangle ,
\label{eq:ces}
\end{equation}
which are obtained by applying excitation operators $\hat{C}_{J}$ to the correlated ground state $|\Psi_{0} \rangle$. The operator form depends on the propagator and the associated property. For the polarization propagator (PP), these operators generate particle–hole excitations that yield excitation energies, whereas for the electron propagator, they describe ionization or electron-attachment processes leading to IP or EA:
\begin{align}
EE: \hspace{1cm}{\hat{C}_{J}} &= {\hat{a}_a^{\dagger}\hat{a}_i; \hat{a}_a^{\dagger}\hat{a}_i\hat{a}_b^{\dagger}\hat{a}j; a < b, i<j; ...} \\
IP: \hspace{1cm}{\hat{C}_{J}} &= {\hat{a}_i; \hat{a}_a^{\dagger}\hat{a}_i\hat{a}_j; i<j; ...}\\
EA: \hspace{1cm}{\hat{C}_{J}} &= {\hat{a}_a^{\dagger}; \hat{a}_a^{\dagger}\hat{a}_b^{\dagger}\hat{a}j; a < b; ...} 
\end{align}
where the occupied orbitals are denoted by $i,j,\dots$, virtual orbitals by $a,b,\dots$, and $p,q,\dots$ denote general indices. Excitations can be organized into $\mu$-hole–$\mu$-particle ($\mu h$–$\mu p$) classes, with $\mu$ counting the number of creation and annihilation operators; an analogous classification applies in the IP and EA cases. Each excitation $J$ is assigned a class $[J]$, with the ground state belonging to the zeroth class, $[J]=0$.

%%%%%%%%%%%%%%% Shorter version %%%%%%%%%%%%%%%%%
The correlated excited states in Eq.~\eqref{eq:ces} are generally nonorthogonal.\cite{mertins1996algebraic}  
A two-step orthogonalization procedure is employed to orthogonalize these states.\cite{schirmer1991closed} In the first step, Gram–Schmidt (GS) orthogonalization is applied to the CES to generate precursor states.
\begin{equation}
|\Psi_J^\#\rangle = |\Psi_J^0\rangle - \sum_{K\,[K]<[J]} |\tilde{\Psi}_K\rangle \langle \tilde{\Psi}_K | \Psi_J^0 \rangle ,
\end{equation}
The resulting precursor states are then made mutually orthonormal by symmetric orthogonalization,
\begin{equation}
|\tilde{\Psi}_I\rangle = \sum_{J\,[J]=[I]} |\Psi_J^\#\rangle S_{IJ}^{-1/2},
\end{equation}
where $S_{IJ} = \langle \Psi_I^\# | \Psi_J^\# \rangle$ is the overlap matrix. 
The resulting intermediate states form a complete orthonormal basis for constructing the ISR secular matrix $\mathbf{M}$.
%%%%%%%%%%%%%%%%%%%%%%%%%%%%%%%%

\begin{equation}
M_{IJ} = \langle \tilde{\Psi}_{I}| \hat{H}^{R}-E_0|\tilde{\Psi}_{J}\rangle ,
\end{equation}
which represents the shifted Hamiltonian $\hat{H}^{4c}-E_0$ in this intermediate-state basis, where $\hat{H}_{R}$ is the relativistic Hamiltonian considered. In this work, $\hat{H}_{R}$ is considered as a 4c-DC or X2CAMF Hamiltonian.
\begin{equation}
    \hat{H}^{\text{4c}} = \sum_{pq}{h^{\text{4c}}_{pq}\hat{a}_p^{\dagger}\hat{a}_q} 
    + \sum_{pqrs}{\frac{1}{4}g^{\text{4c}}_{pqrs}\hat{a}_p^{\dagger}\hat{a}_q^{\dagger}\hat{a}_s\hat{a}_r}
\end{equation}
The exact eigenstates $|\Psi_{n}^{N}\rangle$ can then be expressed as linear combinations of the intermediate states,
\begin{equation}
|\Psi_{n}^{N}\rangle = \sum_{\substack{J}}X_{In}|\tilde{\Psi}_{I}\rangle ,
\end{equation}
where the coefficients $X_{In}$ are obtained from the eigenvectors of $\mathbf{M}$. In compact matrix notation, the secular equation within the ISR takes the form
\begin{equation}
\mathbf{M}\mathbf{X} = \mathbf{X}\boldsymbol{\Omega}, \hspace{1mm} \mathbf{X}^\dagger \mathbf{X} = \mathbf{1},
\label{eq10}
\end{equation}
where $\boldsymbol{\Omega}$ contains the excitation energies on the diagonal.
The secular matrix in Eq.~\eqref{eq10} can be expanded as a perturbative series. By truncating the expansion at a chosen perturbation order ( n ), one obtains the ADC(n) working equations:
\begin{equation}
\label{eq11}
\mathbf{M} = \mathbf{M}^{(0)} + \mathbf{M}^{(1)} + \mathbf{M}^{(2)} + \mathbf{M}^{(3)} + \cdots
\end{equation}
For instance, restricting the series to second order yields the ADC(2) approximation, whereas including terms up to third order results in ADC(3). Dreuw and co-workers~\cite{bauerExploringAccuracyUsefulness2022} introduced a mixed-order variant, known as the scaled matrix ADC(sm-ADC) approach, in which the third-order matrix contribution is scaled by an ad hoc coefficient ( x ):
\begin{equation}
\mathbf{M}_{\text{sm-ADC}[(2)+x(3)]} = \mathbf{M}^{(0)} + \mathbf{M}^{(1)} + \mathbf{M}^{(2)} + x\mathbf{M}^{(3)}
\end{equation}
The scaling constant ( $x$ ) can be varied between 0 and 1. Choosing $x$ = 0 recovers ADC(2), whereas $x$ = 1 corresponds to ADC(3). Based on our previous studies,\cite{mukhopadhyay2025reducedEA,chakraborty2025relativistic} a value of $x$ = 0.5 is used in the present study.

\subsection{Properties in ISR formulation}

The ISR framework described above can be naturally extended to operators other than the Hamiltonian. Let us consider a generic one-particle Hermitian operator $\hat{D}$, which corresponds to a physical observable of interest. In the language of second quantization, such an operator can be written as

\begin{equation}
\hat{D} = \sum_{pq} d_{pq}  \left\{ c_p^\dagger c_q \right\},
\label{eq:D_op}
\end{equation}

Here, $d_{pq}$ represent the matrix elements of the operator $\hat{D}$ expressed in the spinor basis $\{\chi_p\}$,
\begin{equation}
d_{pq} = \langle \chi_p | \hat{D} | \chi_q \rangle,
\label{eq:D_me}
\end{equation}
and the notation with curly braces signifies normal ordering with respect to the reference determinant.

To determine a first-order property of a specific excited state $| \Psi_K^{\mathrm{ex}} \rangle$, corresponding to the operator $\hat{D}$, one needs to evaluate its expectation value, which is given by

\begin{equation}
D_{K} = \langle \Psi_K^{\mathrm{ex}} | \hat{D} | \Psi_K^{\mathrm{ex}} \rangle .
\label{eq:Dk_def}
\end{equation}

In practice, the contributions from the ground and excited states are treated separately. Within the ISR framework, the expectation value of the operator shifted by the ground-state contribution, $(\hat{D}-D_0)$, can be written as

\begin{equation}
\bar{D}_K = \sum_{IJ} Y_{IK}^\ast \langle \tilde{\Psi}_I | \hat{D} - D_0 | \tilde{\Psi}_J \rangle Y_{JK}
= Y_K^\ast G Y_K
\label{eq:D_shifted}
\end{equation}
where the elements of the matrix $\mathbf{G}$ are defined by

\begin{equation}
G_{I J} = \langle \tilde{\Psi}_I | \hat{D} - D_0 | \tilde{\Psi}_J \rangle .
\label{eq:G_matrix}
\end{equation}

These matrix elements are often referred to as modified excited-state transition moments.

To obtain the complete expectation value of the $K^{th}$ excited state, the ground-state contribution,

\begin{equation}
D_0 = \langle \Psi_0 | \hat{D} | \Psi_0 \rangle ,
\label{eq:D0_def}
\end{equation}
must be added to the expectation value of the shifted operator $(\hat{D}-D_0)$ given in Eq.~\eqref{eq:D_shifted}. Accordingly, the total expectation value of the $K^{th}$ excited state can be expressed as
\begin{align}
\bar{D}_K &= \sum_{IJ} Y_{I K}^\ast \langle \tilde{\Psi}_I | \hat{D} - D_0 | \tilde{\Psi}_J \rangle Y_{JK} +
\sum_{IJ} Y_{I K}^\ast \delta_{I J} D_0 Y_{J K} \\
  &= Y_K^\ast G Y_K + D_0 .
  \label{eq:D_shifted_total}
\end{align}

where $D_0$ represents the ground-state component of the total excited-state
property
\begin{equation}
D_0 = \mathrm{Tr}(\rho d)
\end{equation}
Here, $\rho$ denotes the ground-state one-particle reduced density matrix, while $d$ represents the matrix elements of the corresponding one-particle operator. To evaluate the ground-state expectation value at the ADC(2) level, the density matrix must be expanded to include contributions up to second order in perturbation theory, i.e., $\rho = \rho^{(0)} + \rho^{(2)}$.\cite{schirmer2004intermediate}

In addition to first-order properties of excited states, a key quantity of interest is the spectral intensity, which requires evaluation of transition properties between two states. The transition property from the ground state to the K$^{th}$ excited state is defined as

\begin{equation}
T_{0 \to K} = \langle \Psi_K | \hat{D} - D_0 | \Psi_0 \rangle
\end{equation}

Similarly, within the ISR formalism, it can be expressed as,

\begin{equation}
T_{0 \to K} = \sum_I Y_{IK}^\ast F_I % = Y_K^\ast \mathbf{F}
\end{equation}

where

\begin{equation}
F_I = \langle \tilde{\Psi}_I | \hat{D} | \Psi_0 \rangle
\end{equation}

The $\mathbf{F}$ and $\mathbf{G}$ can be expanded in terms of the second quantized operators as,

\begin{equation}
F_I = \sum_{pq} f^{I}_{pq} d_{pq}, 
\qquad 
G_{IJ} = \sum_{pq} g^{IJ}_{pq} d_{pq}
\end{equation}

where,
\begin{equation}
f^{I}_{pq} = \langle \tilde{\Psi}_I | c_p^\dagger c_q | \Psi_0 \rangle, 
\qquad
g^{IJ}_{pq} = \langle \tilde{\Psi}_I | c_p^\dagger c_q | \tilde{\Psi}_J \rangle
\end{equation}

The one-particle excited-state reduced density matrix ($\rho^K_{pq}$) can be written as

\begin{equation}
\rho^K_{pq} = \langle \Psi_K^{\mathrm{ex}} | c_p^\dagger c_q | \Psi_K^{\mathrm{ex}} \rangle
= \sum_{IJ} Y_{IK}^\ast \langle \tilde{\Psi}_I | c_p^\dagger c_q | \tilde{\Psi}_J \rangle Y_{JK}
= Y_K^\ast g_{pq} Y_K
\end{equation}

and the ground-to-excited state transition density matrix ($\rho^{0 \to K}_{pq}$) as

\begin{equation}
\rho^{0 \to K}_{pq} = \langle \Psi_K^{\mathrm{ex}} | c_p^\dagger c_q | \Psi_0 \rangle
= \sum_I Y_{IK}^\ast \langle \tilde{\Psi}_I | c_p^\dagger c_q | \Psi_0 \rangle
= Y_K^\ast f_{pq}
\end{equation}

By contracting these one-particle reduced density matrices with the corresponding one-particle operator, the properties of the excited states and the transition properties can be computed as

\begin{equation}
D_K = \langle \Psi_K^{\mathrm{ex}} | \hat{D} | \Psi_K^{\mathrm{ex}} \rangle
= \sum_{pq} d_{pq} \, \rho^K_{pq}
\end{equation}

\begin{equation}
T_{0 \to K} = \langle \Psi_K^{\mathrm{ex}} | \hat{D} | \Psi_0 \rangle
= \sum_{pq} d_{pq} \, \rho^{0 \to K}_{pq}
\end{equation}

%\textcolor{red}{Cross-check the above equation}
\subsection{The X2CAMF approximation }
\label{sec2.4}

Under the spin-separation scheme,\cite{dyallExactSeparationSpinfree1994}, the two-electron part of the 4c-DC Hamiltonian ($\hat{H}^{\text{4c}}$) can be split into spin-free (SF) and spin-dependent (SD) contributions:
\begin{equation}
\label{eqn33}
    \hat{H}^{\text{4c}} = \sum_{pq}{h^{\text{4c}}_{pq}\hat{a}_p^{\dagger}\hat{a}_q}
    + \frac{1}{4}\sum_{pqrs}{g^{\text{4c,SF}}_{pqrs}\hat{a}_p^{\dagger}\hat{a}_q^{\dagger}\hat{a}_s\hat{a}_r}
    + \frac{1}{4}\sum_{pqrs}{g^{\text{4c,SD}}_{pqrs}\hat{a}_p^{\dagger}\hat{a}_q^{\dagger}\hat{a}_s\hat{a}_r}
\end{equation}

The spin-dependent part can be approximated using the AMF scheme\cite{hessMeanfieldSpinorbitMethod1996,liuAtomicMeanfieldSpinorbit2018,knechtExactTwocomponentHamiltonians2022, zhangAtomicMeanFieldApproach2022}, which exploits the localized character of spin–orbit interactions:
\begin{equation}
\label{eqn34}
    \frac{1}{4}\sum_{pqrs}{g^{\text{4c,SD}}_{pqrs}\hat{a}_p^{\dagger}\hat{a}_q^{\dagger}\hat{a}_s\hat{a}_r}
    \approx \sum_{pq}{g^{\text{4c,AMF}}_{pq}\hat{a}_p^{\dagger}\hat{a}_q}=\sum_{pqiA}{n_{iA}}g^{\text{4c,SD}}_{p_{iA}q_{iA}}\hat{a}_p^{\dagger}\hat{a}_q
\end{equation}
Here, $A$ denotes individual atoms, $i$ corresponds to occupied spinors on atom $A$, and $n_{iA}$ denotes their occupations. Substituting Eq.~(\ref{eqn34}) into Eq.~(\ref{eqn33}) yields
\begin{equation}
\label{eqn35}
    \hat{H}^{\text{4c}} = \sum_{pq}{h^{\text{4c}}_{pq}\hat{a}_p^{\dagger}\hat{a}_q}
    + \frac{1}{4}\sum_{pqrs}{g^{\text{4c,SF}}_{pqrs}\hat{a}_p^{\dagger}\hat{a}_q^{\dagger}\hat{a}_s\hat{a}_r}
    + \sum_{pq}{g^{\text{4c,AMF}}_{pq}\hat{a}_p^{\dagger}\hat{a}_q}
\end{equation}

The 4c Hamiltonian can then be transformed to a two-component picture via the relations between large- and small-component coefficients (the $X$ matrix) and between large-component and two-component coefficients (the $R$ matrix):
\begin{equation}
\label{eqn36}
    C^S=XC^L
\end{equation}
\begin{equation}
\label{eqn37}
    C^L=RC^{\text{2c}}
\end{equation}
The spin-free two-electron contribution $g^{\text{4c,SF}}$ reduces to the nonrelativistic Coulomb integrals $g^{\text{NR}}$ when the two-electron picture-change (2e-pc) corrections are neglected. The resulting two-component Hamiltonian in the X2CAMF approximation is
\begin{equation}
\label{eqn38}
    \hat{H}^{\text{X2CAMF}} = \sum_{pq}{h^{\text{X2C}}_{pq}\hat{a}_p^{\dagger}\hat{a}_q}
    + \frac{1}{4}\sum_{pqrs}{g^{\text{NR}}_{pqrs}\hat{a}_p^{\dagger}\hat{a}_q^{\dagger}\hat{a}_s\hat{a}_r}
    + \sum_{pq}{g^{\text{2c,AMF}}_{pq}\hat{a}_p^{\dagger}\hat{a}_q}
\end{equation}
which may be compactly expressed in terms of an effective one-electron operator:
\begin{equation}
\label{eqn39}
    \hat{H}^{\text{X2CAMF}} = \sum_{pq}{h^{\text{X2CAMF}}_{pq}\hat{a}_p^{\dagger}\hat{a}_q}
    + \frac{1}{4}\sum_{pqrs}{g^{\text{NR}}_{pqrs}\hat{a}_p^{\dagger}\hat{a}_q^{\dagger}\hat{a}_s\hat{a}_r}
\end{equation}
where $h^{\text{X2CAMF}} = h^{\text{X2C}}+g^{\text{2c,AMF}}$. The major benefit of this Hamiltonian is that it eliminates the need to construct relativistic two-electron integrals.

\subsection{Frozen natural spinors}
\label{sec2.3}

Natural spinors\cite{chamoli2025frozen, yuanAssessingMP2Frozen2022, chamoli2025frozen,majee2024reduced,chakraborty2025lowcost} are the relativistic analogues of the natural orbitals introduced by Löwdin.\cite{lowdinQuantumTheoryManyParticle1955} These are obtained as eigenfunctions of the correlated one-particle reduced density matrix.\cite{chamoliRelativisticReducedDensity2024} In ground-state energy computations, natural spinors are typically obtained from the first-order Møller–Plesset (MP1) wave function. Within the frozen natural spinor (FNS) scheme, only the virtual space is optimized using natural spinors, while the occupied space is retained at the Dirac–Hartree–Fock (DHF) level. After the SCF step, a partial AO–MO integral transformation is carried out to form integrals with two external indices. From this, the virtual–virtual block of the one-particle reduced density matrix at the MP2 level can be obtained as
\begin{equation}
    \label{eqn25}
    D_{ab}=\frac{1}{2}\sum_{cij}^{} {\frac{\langle ac||ij \rangle \hspace{0.1cm}\langle ij||bc \rangle}{\varepsilon_{ij}^{ac} \hspace{0.2cm}\varepsilon_{ij}^{bc}}}
\end{equation}

Diagonalization of this virtual–virtual density block $\left( {{D}_{ab}}\right)$ yields the virtual natural spinors (VNS) as eigenvectors $V$ and their corresponding occupation numbers as eigenvalues $\eta$:
\begin{equation}
\label{eqn26}
    D_{ab}V=V\eta
\end{equation}
Virtual spinors with small occupation numbers can be discarded using a cutoff ${{\eta }_{crit}}$,
\begin{equation}
\label{eqn27}
    \tilde{V}=VT
\end{equation}
where ${{T}_{ij}}={{\delta }_{ij}}$ if ${{\eta }_{i}}\ge {{\eta }_{crit}}$, and ${{T}_{ij}}=0$ otherwise.  
The virtual–virtual block of the Fock matrix is then projected into this truncated natural spinor basis,
\begin{equation}
\label{eqn28}
    \tilde{F}={{\tilde{V}}^{\dagger }}F\tilde{V}
\end{equation}
This transformed Fock matrix $\left(\tilde{F}\right)$ is diagonalized to semicanonicalize the truncated basis
\begin{equation}
\label{eqn29}
    \tilde{F}\tilde{Z}=\tilde{Z}\tilde{\epsilon }
\end{equation}
The eigenvectors $\tilde{Z}$, together with the retained natural spinors $\tilde{V}$, form the overall transformation matrix $\left( B \right)$ linking the canonical and truncated FNS bases
\begin{equation}
\label{eqn30}
    B=\tilde{V}\tilde{Z}
\end{equation}
In the subsequent discussion, this procedure for constructing frozen natural spinors from MP2 densities will be referred to as the FNS scheme. While MP2-based natural spinors\cite{chamoli2025frozen, yuanAssessingMP2Frozen2022, chamoli2025frozen, majee2024reduced} are tailored to efficiently recover ground-state correlation energy, as well as ionization potentials\cite{surjuse2022low, SomeshIP} and double ionization potentials, \cite{mandal2025third,mukhopadhyay2025reducedDIP} they are not well suited for excited states\cite{mukhopadhyay2025reduced}. Even in the non-relativistic domain, there is no general consensus on defining optimal natural orbitals for excited states. Recently, we demonstrated that ADC(2) provides an accurate first-order description of electron-attached\cite{mukhopadhyay2025reducedEA}and excited-state wave functions,\cite{mukhopadhyay2025reduced} and that the use of ADC(2)-based natural orbitals yields consistent accuracy across valence, Rydberg, and charge-transfer excited states in the EOM-CCSD method.\cite{manna2025reducedcostequationmotion}

The IP-ADC(3) calculations in this manuscript are performed using standard MP2-based FNS, whereas the EA- and EE-ADC(3) calculations are performed using state-specific frozen natural spinors (SS-FNS). The
state-specific one-particle density matrices are calculated at the relativistic ADC(2) level.\cite{pernpointnerRelativisticPolarizationPropagator2014, pernpointnerFourComponentPolarizationPropagator2018, chakraborty2025relativistic}  For the $k^{th}$ electron attached or excited state, the virtual–virtual block of the density takes the form
\begin{equation}
\label{eqn31}
    D_{ab}^{\text{SS}}(k)=D_{ab}^{\text{MP2}}+D_{ab}^{\text{EA/EE-ADC(2)}}(k)
\end{equation}
where $D_{ab}^{\text{MP2}}$ represents the MP2-level virtual–virtual density block and $D_{ab}^{\text{EA/EE-ADC(2)}}(k)$ is the electron-attached/excited-state contribution from EA/EE-ADC(2). The zeroth-order intermediate-state representation\cite{schirmer2004intermediate} with ADC(2) eigenvectors was used to obtain excited-state densities. The programmable expressions for the one body reduced density matrices used in FNS or SS-FNS scheme for IP, EA and EE are are provided in the Supporting Information. Inserting Eq.~(\ref{eqn31}) into Eq.~(\ref{eqn26}) and following Eqs.~(\ref{eqn26}–\ref{eqn30}) produces the SS-FNS basis for each excited state, in which the ADC(3) calculations are subsequently performed. A schematic representation of the algorithm used for FNS-CD-IP-ADC(3) and SS-FNS-CD-EA/EE-ADC(3) is presented in Figure \ref{fig:algo}.

An energy correction for the FNS truncation can be included by comparing EA- and EE-ADC(2) energies in the canonical basis and the truncated SS-FNS basis. This correction is added to the uncorrected SS-FNS-ADC(3) electron affinity or excitation energy as
{\small
\begin{equation}
\label{eqn32}
        \omega_{\text{SS-FNS-ADC(3)}}^{\text{corrected}}(k) = \omega_{\text{SS-FNS-ADC(3)}}^{\text{uncorrected}}(k) + 
        \omega_{\text{ADC(2)}}^{\text{canonical}}(k)-\omega_{\text{ADC(2)}}^{\text{SS-FNS}}(k)
\end{equation}}

To ensure that each root is solved in its corresponding SS-FNS basis, a root-specific Davidson solver\cite{hiraoGeneralizationDavidsonsMethod1982} is employed. The canonical ADC(2) eigenvectors are transformed to the corresponding SS-FNS basis and act as starting guesses for the subsequent ADC(2) and ADC(3) calculations in the truncated SS-FNS basis.
The FNS truncation correction has not been considered for the FNS-IP-ADC(3) method, as it has been shown that this correction does not lead to any appreciable improvement.\cite{mukhopadhyay2023state,manna2024efficient}.
In relativistic ADC(2)\cite{pernpointnerRelativisticPolarizationPropagator2014, pernpointnerFourComponentPolarizationPropagator2018, chakraborty2025relativistic} calculations targeting electron-attached and excited states, one must additionally handle integrals with one, two, and three external indices in the canonical basis, beyond the two-external integrals required in standard FNS.  In the four-component Dirac–Coulomb (4c-DC) EA-and EE-ADC(2)framework, three external integrals become a major bottleneck for large basis sets. Moreover, the full transformation of integrals—even in the truncated basis—must be repeated for every excited state, making this approach impractical for a conventional  implementation. Therefore, to achieve computational feasibility, additional approximations such as Cholesky decomposition (CD) of two electron integrals must be adopted.
\subsection{Cholesky decomposition of two electron integrals}
\label{sec2.5} 

In the present work, the CD approach\cite{aquilante2011cholesky} is used to calculate two-electron integrals. Unlike the resolution-of-identity (RI) methods,\cite{hattigStructureOptimizationsExcited2005,hohensteinDensityFittingCholesky2010} CD does not rely on a pre-defined auxiliary basis set. Within the CD framework, the electron-repulsion integrals (ERIs) are factorized as
\begin{equation}
\label{eqn41}
    \left(\mu\nu|\kappa\lambda\right) \approx \sum_{P}^{n_{\text{CD}}}{L_{\mu\nu}^{P}L_{\kappa\lambda}^{P}}
\end{equation}
where $\mu, \nu, \kappa, \lambda$ are atomic spinor indices, $L_{\mu\nu}^{P}$ are Cholesky vectors, and $n_{\text{CD}}$ is their rank.
In the one-step procedure adopted here, Cholesky vectors are generated iteratively by selecting the largest remaining diagonal elements of the ERI matrix until this value falls below a preset threshold ($\tau$). The resulting vectors are then transformed into the molecular spinor basis:
\begin{equation}
\label{eqn42}
    L_{pq}^{P} = \sum_{\mu\nu}C_{\mu p}^* L_{\mu\nu}^{P}C_{\nu q}
\end{equation}
From these molecular Cholesky vectors, antisymmetrized integrals are constructed as
\begin{equation}
\label{eqn43}
    \left\langle pq||rs\right\rangle = \sum_{P}^{n_{\text{CD}}} \left ( L_{pr}^{P}L_{qs}^{P} - L_{ps}^{P}L_{qr}^{P} \right )
\end{equation}
Although we use the one-step procedure in the present work, it is worth noting that an efficient two-step algorithm for the Cholesky decomposition of the electron-repulsion integrals has also been developed, in which only the elements of the Cholesky basis are generated during the pivoting procedure.\cite{aquilante2011cholesky,folkestad2019efficient,zhang2021toward} In the present implementation, four-external ($\left\langle ab||cd\right\rangle$) and three-external ($\left\langle ab||ci\right\rangle$) integrals are neither precomputed nor stored but are generated on the fly. All other integrals are explicitly constructed and stored in the truncated FNS basis. More details regarding the FNS and CD-based X2CAMF implementation of relativistic coupled-cluster methods are available in Ref. \cite{chamoli2025frozen}.

The FNS-IP-ADC(3) and SS-FNS-EA/EE-ADC(3) methods, based on the X2CAMF Hamiltonian, have been implemented in our in-house software package, BAGH. \cite{dutta2025bagh} BAGH is written in Python, with Cython and Fortran used to optimize computationally demanding components. It is interfaced with PySCF,\cite{sun2015libcint,sun2018pyscf,sun2020recent} socutils,\cite{socutils} DIRAC,\cite{bast2023dirac23} and GAMESS-US.\cite{barca2020recent} All calculations presented here were performed using the PySCF and socutils interfaces within BAGH.
\begin{figure}[h]
    \centering
    \includegraphics[width=0.9\textwidth]{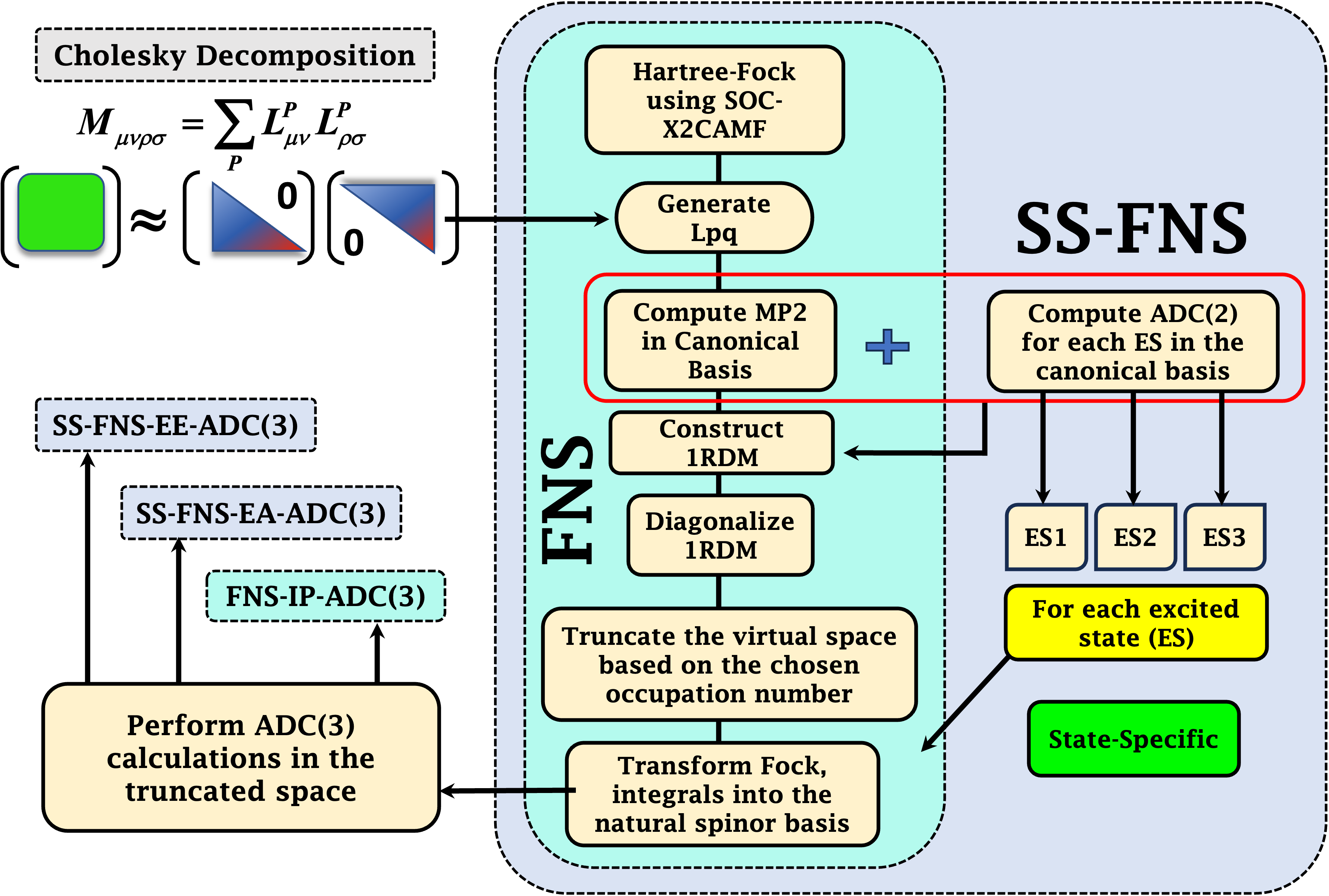} % Replace with your actual path
    % Main caption for the figure
    \caption{\label{fig:algo}The schematic flowchart of CD based FNS-IP-ADC(3), SS-FNS-EE-ADC(3) and SS-FNS-EA-ADC(3) using X2CAMF Hamiltonian.}
    \label{fig:algo}
\end{figure}

\section{Results and discussion}

\subsection{Ionization potential}

\subsubsection{Choice of truncation threshold}

\begin{figure}[h]
    \centering
    \includegraphics[width=0.9\textwidth]{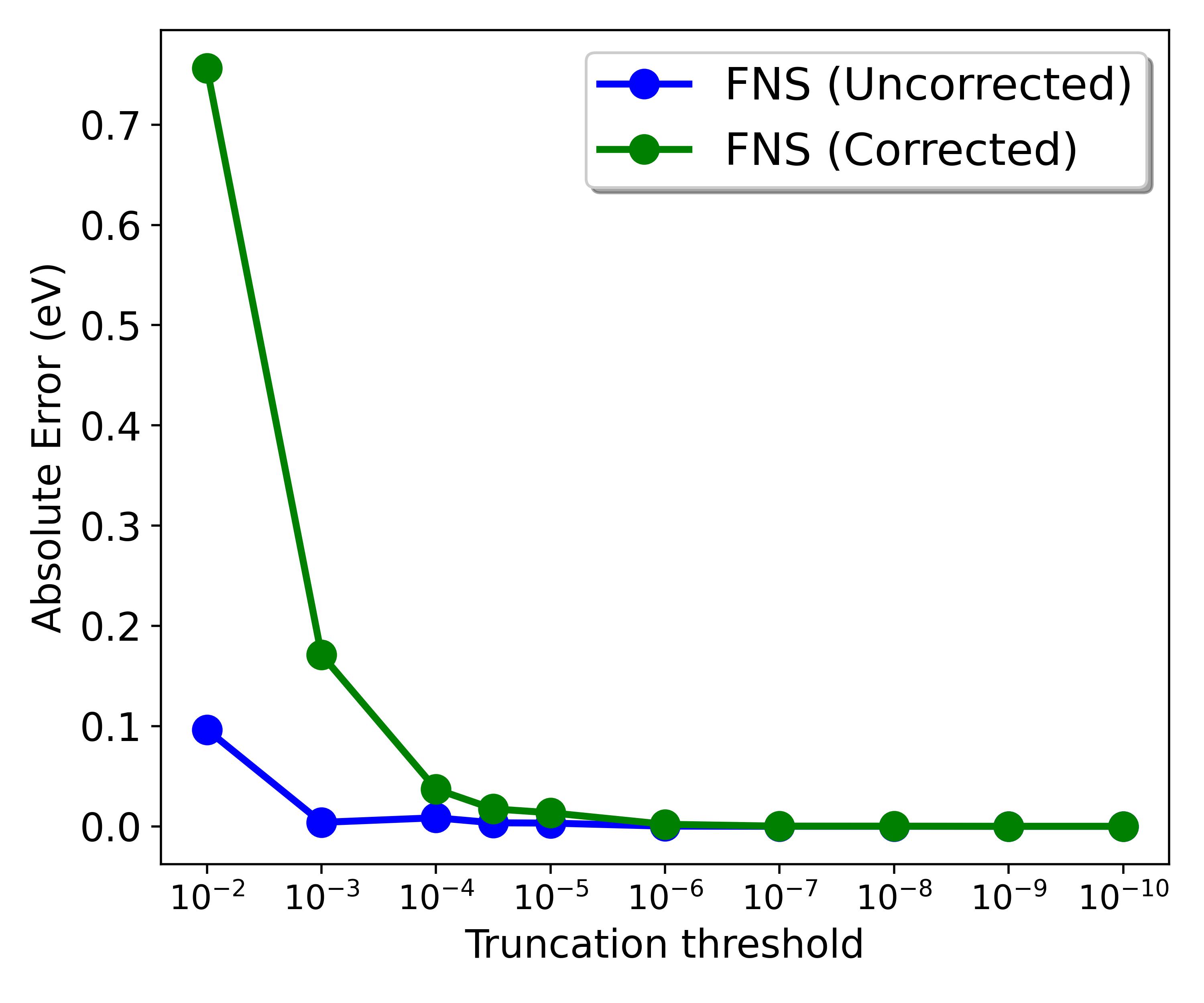} % Replace with your actual path
    % Main caption for the figure
    \caption{\label{fig:ip_thresh_conv}The comparison of absolute error in IP (in eV) for FNS truncation scheme of CD-based X2CAMF version of IP-ADC(3) with respect to the canonical result for IBr molecule using dyall.v2z basis set.}
    \label{fig:IPThresh}
\end{figure}

The accuracy of the FNS-based method depends on the occupation threshold used for truncation. To determine an appropriate FNS threshold, we studied the convergence of IP values with respect to the FNS threshold for the first ionization potential of IBr (See Figure \ref{fig:IPThresh}). The uncontracted dyall.v2z basis set was used for the calculations. The CD threshold was set to $10^{-3}$, following Ref. \citenum{chamoli2025reduced}. The results are nearly converged at a threshold of $10^{-4.5}$. The error are also small at the threshold of $10^{-4}$ and offers enhanced computational efficiency. Therefore, a CD threshold of 
$10^{-3}$ and  an FNS threshold of $10^{-4}$ were used for subsequent calculations, consistent with the LOOSEFNS threshold reported in Ref.\citenum{chamoli2025reduced}. 
\begin{figure}[h]
    \centering
    \includegraphics[width=0.7\textwidth]{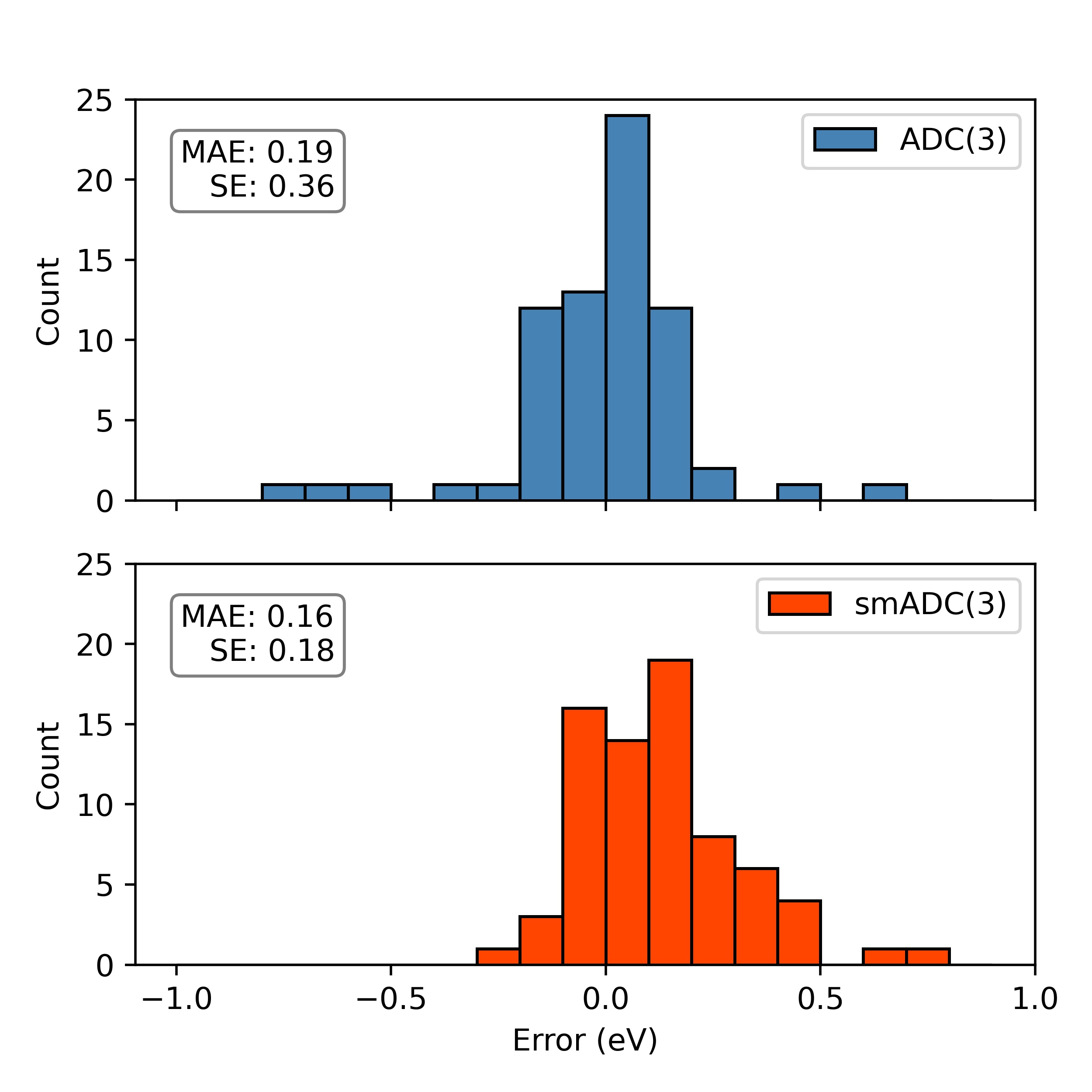} % Replace with your actual path
    % Main caption for the figure
    \caption{\label{fig:soc81-histo} Histogram of errors with respect to experimental value for IP of molecules in SOC-81 dataset with standard FNS-CD-IP-ADC(3) (top) and semiemperically scaled FNS-CD-IP-smADC[(2)+x(3)] (bottom) where $x=0.5$.}
    \label{fig:auh_thresh_conv}
\end{figure}
\subsubsection{Performance on a standard test set}
To further investigate the accuracy of the FNS-IP-ADC(3) method, we calculated the vertical IP of the molecules contained in the SOC-81 dataset\cite{soc81}. The FNS-CD-IP-ADC(3) calculations were performed using the dyall.v4z basis set, and the mean absolute error (MAE), mean error (ME), and standard deviation of errors (SE) relative to the corresponding experimental IP were analyzed. Recently, Zgid and co-workers\cite{abraham2024relativistic} developed a relativistic variant of the fully self-consistent GW (scGW) method, incorporating spin–orbit coupling through the exact two-component (X2C) Hamiltonian. 
\begin{table}[h]
\centering
\caption{\label{table:soc81_error}Comparison of MAE, SE and ME of vertical IP of SOC-81 set using FNS-CD-IP-ADC(3) and previously reported methods with respect to experimental value. All values are in eV}
%\begin{adjustbox}{width=\textwidth}
\vspace{0.5em}
\begin{tabular}{c c c c c c c c c c}
\hline\hline
Method &&&MAE &&&SE &&&ME \\

\hline
$G_{0}W_{0}@$ PBE        &&&0.32 &&&0.23 &&&-0.31 \\
$G_{0}W_{0}@$ PBE0       &&&0.14 &&&0.20 &&&0.00 \\
$scGW$                   &&&0.21 &&&0.17 &&&-0.17 \\
FNS-CD-IP-ADC(3)         &&&0.19 &&&0.36 &&&-0.07 \\
FNS-CD-IP-[ADC(2)+x(3)]  &&&0.16 &&&0.18 &&&0.12\\
FNS-CD-X2CAMF-IP-EOM-CCSD\cite{chamoli2025reduced} &&&0.13 &&&0.18 &&&-0.02 \\
\hline\hline
\end{tabular}
%\end{adjustbox}
\end{table}
They applied both the scGW and $G_{0}W_{0}$ approximation to the SOC-81 dataset, which comprises closed-shell molecules containing heavy elements, and benchmarked their results against experimental first IP. To facilitate a consistent comparison with the GW-based study by Zgid and co-workers, we adopted the same set of 74 molecules from the original SOC-81 dataset used in their benchmarking. In line with their protocol, we excluded seven molecules from the full set - MgI$_{2}$, AsF$_{5}$, AlBr$_{6}$, (C$_{5}$H$_{5}$)$_{2}$Ru, Mo(CO)$_{6}$, TiI$_{4}$, and ZrI$_{4}$ from the statistical analysis to ensure a one-to-one correspondence in the comparison. We also incorporated the same corrections to experimental IP used in analysis for HgCl$_{2}$, KBr, and RbBr, where the originally reported IPs in the NIST database corresponded to the second IPs (due to spin–orbit splitting) rather than the first. The IPs of all molecules, including the excluded systems, are presented in Table S1 of the Supporting Information.

As shown in Table~\ref{table:soc81_error}, the standard FNS-CD-IP-ADC(3) approach yields a MAE of 0.19 eV and a ME of -0.07 eV, indicating a systematic underestimation relative to experimental IPs. However, the SE is 0.36 eV, indicating a relatively larger variance in errors across the dataset. The results are inferior to the corresponding EOM-CCSD values. In contrast, the scaled matrix FNS-CD-IP-[ADC(2)+x(3)] method, in which the third-order ADC contribution is scaled by an empirical factor, $x = 0.5$, exhibits excellent performance and achieves an improved MAE of 0.16 eV and a lower SE of 0.18 eV. The histogram analysis in Figure \ref{fig:soc81-histo} further supports this trend, showing tighter clustering of errors around the experimental benchmark for the FNS-CD-IP-[ADC(2)+x(3)] method. To contextualize these findings, we benchmarked both ADC-based methods against GW-based approaches applied to the same SOC-81 subset. Table~\ref{table:soc81_error} compiles the MAE, SE, and ME obtained from one-shot and self-consistent GW approaches, $G_{0}W_{0}@\mathrm{PBE}$, $G_{0}W_{0}@\mathrm{PBE0}$, and scGW, each implemented with an exact two-component (X2C) relativistic Hamiltonian. The $G_{0}W_{0}@\mathrm{PBE}$ scheme performs the worst (MAE = 0.32 eV, ME = –0.31 eV), whereas $G_{0}W_{0}@\mathrm{PBE0}$ improves accuracy (MAE = 0.14 eV, ME = 0.00 eV, SE = 0.20 eV), highlighting sensitivity of $G_{0}W_{0}$ method to the choice of DFT functional. Fully self-consistent scGW yields an intermediate MAE of 0.21 eV (ME = –0.17 eV, SE = 0.17 eV) but at substantially higher computational cost due to iterative Dyson-equation solutions. The FNS-CD-IP-[ADC(2)+x(3)] method exhibits comparable or superior accuracy relative to GW techniques while retaining significant computational advantages. The low MAE and SE values underscore the efficacy of combining frozen natural spinors with semi-empirical scaling for reliable IP predictions in molecular systems where relativistic effects are significant.

\subsubsection{Comparison with four-component results}

\begin{table*}[h!]
\centering
\caption{Spin-orbit splitting (eV) of halogen oxides calculated with different methods.}
\label{tab:sos_halogen_oxides}
\begin{tabular}{l l c c c c c c}
\hline
\hline
Molecule & Method & $^{2}\Pi_{3/2}$ & $N_{v,\text{act}}$ & $^{2}\Pi_{1/2}$ & Splitting & Expt & Error$^c$ \\
\hline
ClO$^-$ & 4c-IP-ADC(3)$^a$ & 2.4158 &      & 2.4659 & 0.0502 & 0.0397 & 0.0105 \\
        & 4c-IP-ADC[(2)+x(3)]$^a$ & 1.7446 &      & 1.7862 & 0.0416 &        & 0.0019 \\
        & 4c-IP-EOM-CCSD$^a$ &       &      &       & 0.0422 &        & 0.0025 \\
        & FNS-CD-IP-ADC(3)$^a$ & 2.3367 & 106  & 2.3862 & 0.0495 &        & 0.0098 \\
        & FNS-CD-IP-ADC[(2)+x(3)]$^a$ & 1.7087 & 106  & 1.7499 & 0.0412 &        & 0.0015 \\
        & FNS-CD-IP-ADC(3)$^b$ & 2.3949 & 112  & 2.4476 & 0.0527 &        & 0.0130 \\
        & FNS-CD-IP-ADC[(2)+x(3)]$^b$ & 1.7687 & 112  & 1.8090 & 0.0403 &        & 0.0006 \\
\hline
BrO$^-$ & 4c-IP-ADC(3)$^a$ & 2.5819 &      & 2.7470 & 0.1651 & 0.1270 & 0.0381 \\
        & 4c-IP-ADC[(2)+x(3)]$^a$ & 1.9125 &      & 2.0347 & 0.1222 &        & 0.0048 \\
        & 4c-IP-EOM-CCSD$^a$ &       &      &       & 0.1294 &        & 0.0024 \\
        & FNS-CD-IP-ADC(3)$^a$ & 2.5305 & 132  & 2.6925 & 0.1620 &        & 0.0350 \\
        & FNS-CD-IP-ADC[(2)+x(3)]$^a$ & 1.8923 & 132  & 2.0130 & 0.1207 &        & 0.0063 \\
        & FNS-CD-IP-ADC(3)$^b$ & 2.5347 & 154  & 2.7004 & 0.1657 &        & 0.0387 \\
        & FNS-CD-IP-ADC[(2)+x(3)]$^b$ & 1.9337 & 154  & 2.0530 & 0.1193 &        & 0.0077 \\
\hline
IO$^-$  & 4c-IP-ADC(3)$^a$ & 2.5889 &      & 2.9760 & 0.3871 & 0.2593 & 0.1278 \\
        & 4c-IP-ADC[(2)+x(3)]$^a$ & 2.0161 &      & 2.2791 & 0.2630 &        & 0.0037 \\
        & 4c-IP-EOM-CCSD$^a$ &       &      &       & 0.2776 &        & 0.0183 \\
        & FNS-CD-IP-ADC(3)$^a$ & 2.5880 & 140  & 2.9674 & 0.3794 &        & 0.1201 \\
        & FNS-CD-IP-ADC[(2)+x(3)]$^a$ & 2.0231 & 140  & 2.2792 & 0.2561 &        & 0.0032 \\
        & FNS-CD-IP-ADC(3)$^b$ & 2.5184 & 172  & 2.9171 & 0.3987 &        & 0.1394 \\
        & FNS-CD-IP-ADC[(2)+x(3)]$^b$ & 2.0683 & 172  & 2.3235 & 0.2552 &        & 0.0041 \\
\hline
\hline
\end{tabular}

\begin{flushleft}
$^a$ X: dyall.acv3z, O: uncontracted aug-cc-pVTZ. \\
$^b$ X: dyall.acv4z, O: uncontracted aug-cc-pVQZ.\\
$^c$ Error with respect to the experiment.
\end{flushleft}
\end{table*}

Table \ref{tab:sos_halogen_oxides} presents the spin–orbit splittings of the halogen oxide anions XO$^-$ (X=Cl, Br, I) obtained with various relativistic electron detachment methods, and compares them against available experimental and previously reported theoretical values. The focus here is on the performance of the FNS-CD variants of IP-ADC in relation to the conventional 4c implementations. The spin--orbit splittings of halogen monoxide anions XO$^{-}$ were calculated using the Dyall.acv3z basis set for the halogen atom (X) and an uncontracted aug-cc-pVTZ basis set for the oxygen atom. 
The results in Table~\ref{tab:sos_halogen_oxides} show that the FNS-CD-IP-ADC(3) methods based on X2CAMF Hamiltonian reproduce the 4c values with very good accuracy while substantially reducing the computational cost through a truncated virtual space. Across ClO$^-$, BrO$^-$, and IO$^-$, the FNS-CD-IP-ADC(3) spin–orbit splittings remain consistently within 0.003–0.005 eV of the corresponding 4c-IP-ADC(3) values, independent of the basis set. A comparable behavior is observed for the semiempirical FNS-CD-IP-ADC[(2)+($x$)(3)] approach, which closely reproduces the 4c results, with deviations in the splittings typically below 0.002 eV. Compared to experiment, both 4c and FNS-CD variants at the standard ADC(3) level tend to deviate more, especially for the heavier IO$^-$, while the semi-emperically scaled treatment significantly improves agreement and yields errors on the order of a few meV. Thus, FNS-CD offers a reliable and computationally more affordable alternative to 4c calculations without sacrificing accuracy in describing spin–orbit splittings.

\subsection{Excitation energy}

\subsubsection{Choice of truncation threshold}
We have selected gold hydride (AuH) as a benchmark system to investigate the dependence of EE on the truncation threshold. Specifically, we focused on the $0^{+}(\text{II})$ excited state, for which experimental EE data are available.  Figure~\ref{fig:auh_thresh_conv} illustrates the variation of the absolute error in EE as a function of the truncation threshold, comparing three computational schemes, namely SS-FNS-CD-EE-ADC(3), perturbatively corrected SS-FNS-CD-EE-ADC(3), and FNS-CD-EE-ADC(3), denoted as SS-FNS (Uncorrected), SS-FNS (Corrected), and FNS, respectively. The singly augmented dyall.v2z basis set was used for Au, and the uncontracted aug-cc-pVDZ basis set is used for H in the calculations. It is evident that the FNS method exhibits a strong dependence on the truncation threshold. At a truncation threshold of $10^{-2}$, the absolute error is large, nearly 8 eV. However, the error decreases with tighter thresholds, reaching below 0.1 eV at a threshold of $10^{-6}$, and eventually plateauing with negligible error as the threshold is further tightened. In contrast, both SS-FNS (Uncorrected) and SS-FNS (Corrected) exhibit consistently low errors across the entire range of truncation thresholds. 
\begin{figure}[h!]
    \centering
    \includegraphics[width=0.9\textwidth]{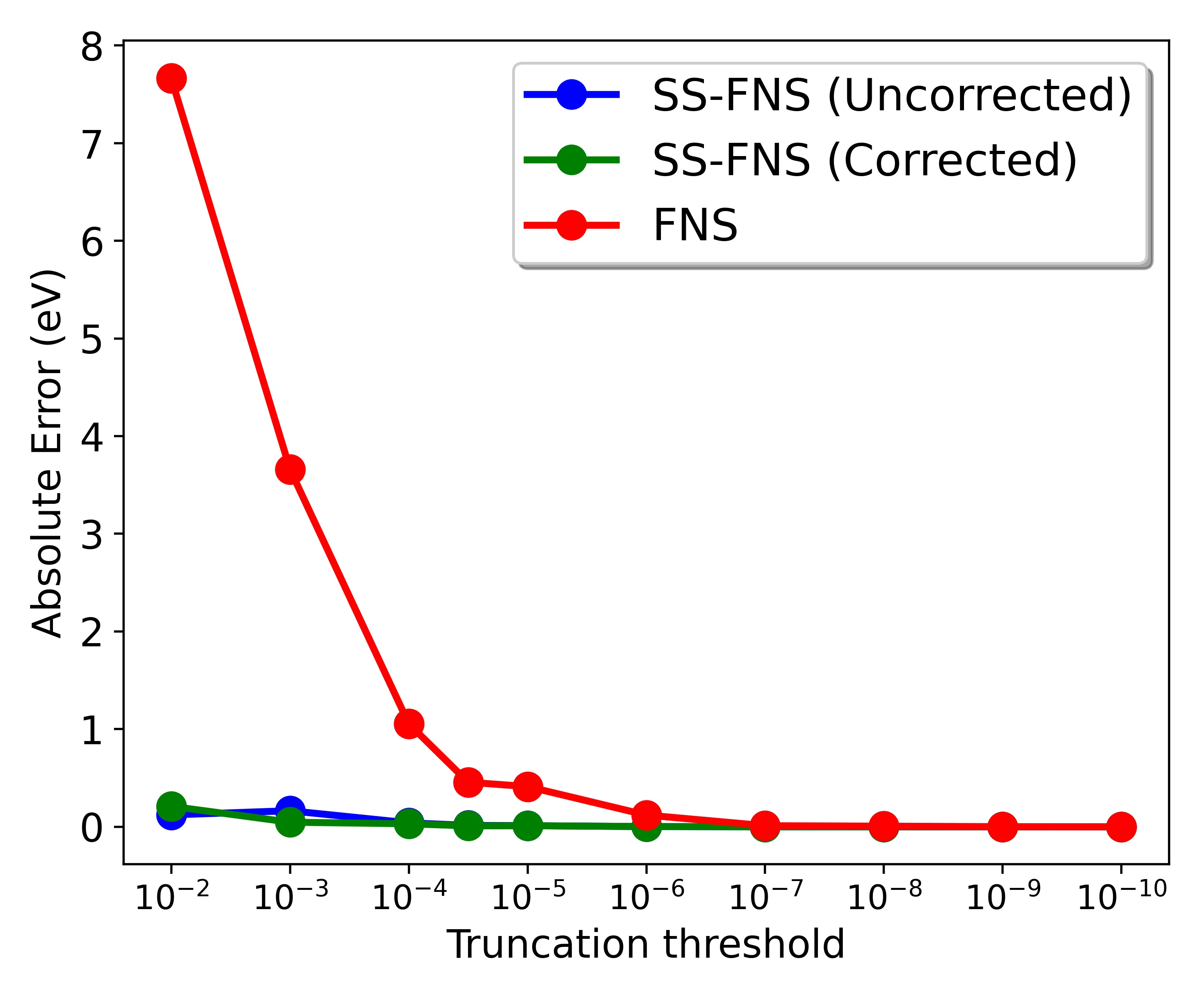} % Replace with your actual path
    % Main caption for the figure
    \caption{\label{fig:epsart}The comparison of absolute error in excitation energies (in eV) for FNS and SS-FNS truncation schemes of CD-based X2CAMF version of EE-ADC(3) with respect to the canonical result for AuH molecule (0$^{+}$(II) state) in singly augmented dyall.v2z basis set for Au and uncontracted aug-cc-pVDZ basis set for H at different truncation thresholds.}
    \label{fig:auh_thresh_conv}
\end{figure}
The SS-FNS method maintains a relatively stable error profile, with errors remaining below 0.2 eV even at a threshold of $10^{-2}$, and gradually decreasing to less than 0.01 eV at tighter thresholds. The inclusion of perturbative corrections for the truncation improves the convergence behavior of the SS-FNS-CD-EE-ADC(3) method, although the effect is small. This confirms that the SS-FNS-CD-EE-ADC(3) scheme is not only accurate but also stable with respect to the choice of truncation threshold.  In contrast, the FNS approach requires a very tight threshold (e.g.,$10^{-6}$ or tighter) to achieve comparable accuracy, making it less practical in such contexts. It is noteworthy that at a truncation threshold of $10^{-4.5}$, the SS-FNS-CD-EE-ADC(3) approach retains 132 out of 382 virtual spinors, corresponding to a truncation of approximately 65\% of the virtual space. Under the same threshold, the FNS-CD-ADC(3) method retains 128 virtual spinors, corresponding to a truncation of 66\%. Despite comparable levels of truncation, the corrected SS-FNS-CD-EE-ADC(3) scheme achieves markedly superior accuracy, exhibiting an absolute error of only 0.0096 eV, in contrast to the significantly larger error of 0.46 eV observed for the FNS-CD-ADC(3) approach.
This result highlights the efficacy of the relativistic SS-FNS-CD-EE-ADC(3) approach in reliably recovering canonical-level accuracy even under aggressive truncation conditions.  A truncation threshold of $10^{-4.5}$ was employed for all subsequent excitation energy calculations across atoms and molecules in this section.

\begin{table}[h]
\centering
\resizebox{\textwidth}{!}{%
\caption{Comparison of excitation energies (in eV) of Ga\textsuperscript{+}, In\textsuperscript{+}, and Tl\textsuperscript{+} from SS-FNS-EE-ADC(3), 4c-EE-ADC(3), and experiment using t-aug-dyall.v3z.}
\vspace{0.5em}
\label{tab:ga_un_tl}
\begin{tabular}{l c c c c c c c}
\hline\hline
 & & 
 \multicolumn{2}{c}{\underline{SS-FNS-EE-ADC(3)}} &
 \multicolumn{2}{c}{\underline{SS-FNS-EE-ADC[(2)+x(3)]}} &
 \underline{4c-EE-ADC(3)} & Expt. \\
 %\cline{3-6}
 & & Uncorrected & Corrected & Uncorrected & Corrected & Canonical & \\
\hline
Ga\textsuperscript{+} & $^1S_0 \rightarrow\ ^3P_0$ & 5.545 & 5.550 & 5.469 & 5.474 & 5.534 & 5.874 \\
 & $^1S_0 \rightarrow\ ^3P_1$ & 5.599 & 5.604 & 5.523 & 5.528 & 5.589 & 5.928 \\
 & $^1S_0 \rightarrow\ ^3P_2$ & 5.713 & 5.718 & 5.632 & 5.639 & 5.702 & 6.044 \\
 & $^1S_0 \rightarrow\ ^1P_1$ & 8.628 & 8.618 & 8.669 & 8.659 & 8.610 & 8.766 \\
 & Splitting & & & & & & \\
 & $^3P_0 \rightarrow\ ^3P_1$ & 0.054 & 0.054 & 0.054 & 0.054 & 0.054 & 0.055 \\
 & $^3P_1 \rightarrow\ ^3P_2$ & 0.114 & 0.114 & 0.109 & 0.111 & 0.113 & 0.116 \\
\\
In\textsuperscript{+} & $^1S_0 \rightarrow\ ^3P_0$ & 4.855 & 4.858 & 4.796 & 4.799 & 4.851 & 5.242 \\
 & $^1S_0 \rightarrow\ ^3P_1$ & 4.986 & 4.989 & 4.926 & 4.930 & 4.982 & 5.375 \\
 & $^1S_0 \rightarrow\ ^3P_2$ & 5.280 & 5.283 & 5.218 & 5.221 & 5.275 & 5.682 \\
 & $^1S_0 \rightarrow\ ^1P_1$ & 7.686 & 7.673 & 7.726 & 7.714 & 7.673 & 7.816 \\
 & Splitting & & & & & & \\
 & $^3P_0 \rightarrow\ ^3P_1$ & 0.131 & 0.131 & 0.130 & 0.131 & 0.131 & 0.133 \\
 & $^3P_1 \rightarrow\ ^3P_2$ & 0.294 & 0.294 & 0.292 & 0.291 & 0.293 & 0.307 \\
\\
Tl\textsuperscript{+} & $^1S_0 \rightarrow\ ^3P_0$ & 5.982 & 5.981 & 6.137 & 6.137 & 5.983 & 6.131 \\
 & $^1S_0 \rightarrow\ ^3P_1$ & 6.345 & 6.342 & 6.498 & 6.496 & 6.344 & 6.496 \\
 & $^1S_0 \rightarrow\ ^3P_2$ & 7.449 & 7.450 & 7.584 & 7.586 & 7.451 & 7.653 \\
 & $^1S_0 \rightarrow\ ^1P_1$ & 9.219 & 9.213 & 9.407 & 9.401 & 9.221 & 9.381 \\
 & Splitting & & & & & & \\
 & $^3P_0 \rightarrow\ ^3P_1$ & 0.363 & 0.362 & 0.361 & 0.359 & 0.361 & 0.365 \\
 & $^3P_1 \rightarrow\ ^3P_2$ & 1.104 & 1.107 & 1.086 & 1.090 & 1.107 & 1.157 \\
\hline\hline
\end{tabular}
}
\end{table}

\subsubsection{Excitation energies and fine-structure splitting of Ga$^{+}$, In$^{+}$, and Tl$^{+}$}

Table ~\ref{tab:ga_un_tl} presents a comparative analysis of the excitation energies (in eV) of group 13 cations Ga$^+$, In$^+$, and Tl$^+$ obtained using the SS-FNS-CD-EE-ADC(3) method, the fully relativistic 4c-EE-ADC(3) method, and experimental values.  Calculations were performed using the triply augmented Dyall valence triple-zeta (t-aug-dyall.v3z) basis set. Virtual spinors with energies up to 2000 a.u. were included in the correlation treatment, and the frozen-core approximation was applied to all atoms. This resulted in 278, 300, and 402 virtual spinors for the canonical calculations of Ga$^+$, In$^+$, and Tl$^+$, respectively. For all three ions, the four low-lying excited states, namely $^3P_0$, $^3P_1$, $^3P_2$, and $^1P_1$, were investigated. For all three atoms, the uncorrected and corrected SS-FNS-CD-EE-ADC(3) EEs are in close agreement, differing only in the third decimal place. Comparison with the fully relativistic 4c-EE-ADC(3) results reveals that the SS-FNS-CD-EE-ADC(3) approach, especially the corrected variant, provides a consistent approximation to the fully relativistic treatment, with deviations typically within  $\sim$ 0.01–0.02 eV. All variants of the ADC(3) method, however, underestimate the excitation energies relative to experiment, which may be attributed to the absence of higher-order correlation and relativistic effects.  For example, in Ga$^+$, the $^1S_0 \rightarrow\ ^3P_0$ transition energy is underestimated by approximately 0.32 eV by the SS-FNS-CD-EE-ADC(3) method when compared to the experimental result. The use of the ADC[(2)+x(3)] approximation does not show any appreciable improvement over the standard ADC(3) method. The calculated fine-structure splittings between the $^3P_J$ states show excellent agreement with the experimental data. For all three cations, both SS-FNS-CD-EE-ADC(3) and 4c-EE-ADC(3) reproduce the energy gaps between the $^3P_0 \rightarrow ^3P_1$ and $^3P_1 \rightarrow ^3P_2$ transitions, with deviations well below 0.01 eV. The largest splitting, observed in Tl$^+$ (1.104–1.107 eV), is also accurately captured, demonstrating the ability of both approaches to account for spin-orbit coupling effects in heavy elements. The use of the larger t-aug-dyall.v4z basis set leads to an improvement in the EE values, but the improvement in fine-structure splittings is negligible (see Table S2 in Supporting Information).
\subsubsection{Excitation energy of I$_{3}^{-}$ molecule}

\begin{table}[h!]
\centering
\resizebox{\textwidth}{!}{%
\caption{
\label{tab:I3-}
\textbf{Comparison of excitation energies (in eV) of I$_{3}^{-}$ ion calculated using SS-EE-ADC(3) and SS-FNS-EE-ADC(3) methods with the dyall.av3z basis set, compared to relativistic 4c-DC-EE-ADC(3) canonical results. The SS-FNS calculations are performed using the X2CAMF Hamiltonian and employing CD.}
}
\renewcommand{\arraystretch}{1.1}
\setlength{\tabcolsep}{5pt}
\begin{tabular}{c c c c c c c c c c c}
\hline
\hline
State & $\Omega$ & \underline{4c-DC-EOM-CCSD}\cite{sheeEquationofmotionCoupledclusterTheory2018}&
\underline{4c-DC-EE-ADC(3)} &
\multicolumn{2}{c}{\underline{SS-FNS-EE-ADC(3)}} &
\multicolumn{2}{c}{\underline{SS-FNS-EE-ADC[(2)+x(3)]}} &
\multicolumn{2}{c}{\underline{SS-FNS-EE-EOM-CCSD}} &
$N_{\mathrm{vir}}$ \\
%\cline{4-5}\cline{6-7}
 & & & & Uncorrected & Corrected & Uncorrected & Corrected &Uncorrected &Corrected & \\
\hline
1  & $2_g$    & 2.24  & 2.06 & 2.07 & 2.07 & 2.21 & 2.21 &2.27 &2.26  &254 \\
2  & $1_g$    & 2.37  & 2.19 & 2.20 & 2.19 & 2.41 & 2.40 &2.41 &2.40  &256 \\
3  & $0_u^-$  & 2.37  & 2.23 & 2.23 & 2.23 & 2.41 & 2.41 &2.41 &2.41  &254 \\
4  & $1_u$    & 2.38  & 2.23 & 2.24 & 2.24 & 2.34 & 2.34 &2.40 &2.40  &254 \\
5  & $0_g^-$  & 2.84  & 2.66 & 2.67 & 2.67 & 2.81 & 2.81 &2.86 &2.86  &254 \\
6  & $0_g^+$  & 2.89  & 2.71 & 2.71 & 2.71 & 2.87 & 2.86 &2.92 &2.91  &254 \\
7  & $1_g$    & 3.07  & 2.87 & 2.88 & 2.88 & 3.03 & 3.03 &3.09 &3.08  &256 \\
8  & $2_u$    & 3.32  & 3.13 & 3.13 & 3.13 & 3.27 & 3.27 &3.35 &3.34  &256 \\
9  & $1_u$    & 3.41  & 3.19 & 3.20 & 3.19 & 3.35 & 3.35 &3.43 &3.42  &256 \\
10 & $0_u^+$  & 3.66  & 3.43 & 3.44 & 3.44 & 3.60 & 3.59 &3.68 &3.67  &262 \\
11 & $2_g$    & 4.09  & 3.88 & 3.89 & 3.89 & 4.06 & 4.06 &4.12 &4.12  &258 \\
12 & $0_u^-$  & 4.08  & 3.89 & 3.90 & 3.90 & 4.19 & 4.18 &4.10 &4.09  &254 \\
13 & $1_u$    & 4.18  & 3.96 & 3.97 & 3.97 & 4.04 & 4.04 &4.20 &4.19  &260 \\
14 & $1_g$    & 4.21  & 3.99 & 3.99 & 3.99 & 4.14 & 4.13 &4.24 &4.23  &256 \\
15 & $0_u^+$  & 4.49  & 4.28 & 4.29 & 4.29 & 4.43 & 4.43 &4.50 &4.49  &266 \\
16 & $0_g^-$  & 4.69  & 4.49 & 4.50 & 4.50 & 4.67 & 4.67 &4.72 &4.71  &256 \\
17 & $0_g^+$  & 4.70  & 4.50 & 4.51 & 4.50 & 4.68 & 4.68 &4.73 &4.72  &256 \\
18 & $1_g$    & 4.90  & 4.85 & 4.89 & 4.87 & 4.94 & 4.92 &4.94 &4.93  &266 \\
\hline
MAD  &    &      &           & 0.009 & 0.007& &  &  \\
STD  &    &      &           & 0.008 & 0.006& &  &  \\
RMSD &    &      &           & 0.013 & 0.009& &  &  \\
\hline
\hline
\end{tabular}
}
\end{table}

Table \ref{tab:I3-} presents a comparison of EEs for the I$_{3}^{-}$ ion, computed using the X2CAMF-SS-FNS-CD-EE-ADC(3) method with the dyall.av3z basis set. These values are benchmarked against 4c-EE-ADC(3) results, which serve as a reference. Following our previous work on relativistic ADC(3), all calculations in this study were performed using the dyall.av3z basis set. Following Ref. \citenum{sheeEquationofmotionCoupledclusterTheory2018}, a total of 52 electrons were correlated, and all spinors with orbital energies in the range of -3.0 to 12.0 a.u. were included, resulting in 332 virtual spinors in the canonical space. For each excited state, both uncorrected and perturbatively corrected excitation energies obtained from the SS-FNS formalism are reported, along with the number of active virtual spinors ($N{_\mathrm{vir}}$) retained in the truncated virtual space. The uncorrected SS-FNS-CD-EE-ADC(3) results show excellent agreement with the 4c-ADC(3) EEs across all 18 excited states considered. The maximum deviation is minimal, and the energy ordering of the excited states remains consistent throughout. The inclusion of the perturbative correction further improves the results, leading to a slight reduction in the overall deviation from the 4c reference values. In many cases, the corrected EEs are numerically indistinguishable from the uncorrected ones, indicating that the dominant contributions are already captured accurately within the SS-FNS space. To quantitatively assess accuracy, we computed the mean absolute deviation (MAD), standard deviation (STD), and root-mean-square deviation (RMSD) relative to the 4c results. The uncorrected SS-FNS-CD-EE-ADC(3) yields a MAD of 0.009 eV, an STD of 0.008 eV, and an RMSD of 0.013 eV. Upon applying the perturbative correction, these values slightly improve to a MAD of 0.007 eV, an STD of 0.006 eV, and an RMSD of 0.009 eV. These small deviations indicate the accuracy and reliability of the SS-FNS-CD-EE-ADC(3) framework in describing relativistic EE with X2CAMF Hamiltonian. In addition to the numerical accuracy, the SS-FNS approach offers significant computational savings by operating in a reduced virtual space (e.g., 254–266 virtual spinors versus 332 in the full canonical basis). The standard SS-FNS-CD-EE-ADC(3) results show significant deviation as compared to the EOM-CCSD values, whereas the SS-FNS-CD-EE-ADC[(2)+x(3)] method shows better agreement with the latter. 

\subsubsection{Transition properties of Xe atom}
Table~\ref{tab:tableXe} reports the EEs and transition dipole moments (TDM) for selected excited states of the xenon atom, calculated using the SS-FNS-EE-ADC(3) and 4c-EE-ADC(3) methods and compared with experimental data. All calculations were performed using the d-aug-Dyall.ae3z basis set, employing the frozen-core approximation and a virtual spinor energy cutoff of 361 a.u. This choice results in a correlation space consisting of 18 occupied and 364 virtual spinors. The same basis set and correlation space were used as in our previous 4c-EE-ADC(3) implementation.\cite{chakraborty2025relativistic} 
The SS-FNS-CD-EE-ADC(3) closely reproduces both the EEs and TDMs obtained from 4c-EE-ADC(3). The deviations between SS-FNS and 4c are within 0.03 eV for EEs and less than 0.005 a.u. for TDMs, indicating that in the SS-FNS framework, the truncated virtual space is sufficient to capture the essential correlation and relativistic effects for these transitions. Interestingly, with a truncation threshold of $10^{-4.5}$, approximately 56\% of the 364 virtual spinors are excluded, leaving only 158 active virtual spinors for the SS-FNS-CD-EE-ADC(3) calculations. 
However, the performance of SS-FNS-CD-EE-ADC(3) is mixed relative to the experimental results. The calculated transition moment shows good agreement with the experimental results except for the  $5p^{5}(^2P{3/2})5d,^2[3/2]_1^o$ states. However, the excitation energy values are significantly underestimated as compared to the experiment for all states. The performance is similar to that of the canonical 4c-ADC(3) method. The excitation energy values can be improved by using the semi-empirically scaled  SS-FNS-CD-EE-ADC[(2)+x(3)] method. However, the agreement with experiment for transition moments significantly deteriorates upon using SS-FNS-CD-EE-ADC[(2)+x(3)]. The trend is consistent with that observed by Dreuw and co-workers ~\cite{bauerExploringAccuracyUsefulness2022} for the non-relativistic case.
%\newpage

\begin{table}[h!]
\centering
\resizebox{\textwidth}{!}{%
\begin{threeparttable}
\caption{\label{tab:tableXe}Comparison of transition dipole moments (TDM in a.u.) for selected excited states of the Xe atom obtained using different variants of ADC(3) using d-aug-dyall.ae3z basis set and compared with experimental results. The SS-FNS calculations are performed using X2CAMF Hamiltonian and employing CD.}
\begin{tabular}{c ccc c ccc c cc c cc}
\hline
\hline
{State} 
& \multicolumn{3}{c}{SS-FNS-EE-ADC(3)}
& \hspace{2pt} 
& \multicolumn{3}{c}{SS-FNS-EE-ADC[(2)+x(3)]}
& \hspace{2pt} 
& \multicolumn{2}{c}{4c-EE-ADC(3)} 
& \hspace{2pt} 
& \multicolumn{2}{c}{Expt.\cite{sym12111845,10.1063/1.1800011}} \\
\cline{2-4} \cline{6-8} \cline{10-11} \cline{13-14}
& EE\tnote{1} & EE\tnote{2} & TDM 
& & EE\tnote{1} & EE\tnote{2} & TDM 
& & EE & TDM & & EE  &TDM\\
\hline
A & 8.20  & 8.19 & 0.646 & & 8.42& 8.41& 0.295& & 8.17 & 0.643 & & 8.43&0.654\\
B & 9.33 &  9.32 & 0.518 & & 9.56& 9.55& 0.258& & 9.31 & 0.522 & & 9.57&0.521\\
C & 9.72 &  9.71 & 0.088 & & 9.97& 9.96& 0.040& & 9.70 & 0.047 & & 9.92&0.120\\
D & 10.44 & 10.43 & 0.901 & & 10.69& 10.68& 0.469& & 10.42 & 0.892 & & 10.40&0.704\\
\hline
\hline
\end{tabular}
\begin{tablenotes}[flushleft]
\footnotesize
\item[1] Uncorrected excitation energy.
\item[2] Corrected excitation energy.\\
A. $5p^{5}({^2P_{3/2}})6s {^2[3/2]_{1}^{o}}$ \\
B. $5p^{5}({^2P_{1/2}})6s {^2[1/2]_{1}^{o}}$ \\
C. $5p^{5}({^2P_{3/2}})5d {^2[1/2]_{1}^{o}}$ \\
D. $5p^{5}({^2P_{3/2}})5d {^2[3/2]_{1}^{o}}$
\end{tablenotes}
\end{threeparttable}
}
\end{table}

\subsection{Electron Attachment}

\subsubsection{Choice of truncation threshold}
Figure \ref{fig:ea_thresh_conv} presents the convergence behavior of the absolute error of the computed EA of the IBr molecule as a function of the truncation threshold in the SS-FNS and FNS truncation schemes, using the CD-based X2CAMF implementation of EA-ADC(3) with the s-aug-dyall.v2z basis set. The FNS approach shows a relatively slow and nonuniform convergence pattern, with large deviations up to $\sim$ 3.5 eV, at lower truncation thresholds. In the FNS scheme, the diffuse orbitals are discarded during truncation because they make a negligible contribution to the ground-state correlation energy. However, these diffuse functions are essential for the EA problem, and discarding them is responsible for the slow convergence of EA values in the FNS-CD-EA-ADC(3) method.  In contrast, both the uncorrected and corrected SS-FNS schemes exhibit much faster and smoother convergence toward the canonical result. Even at a threshold of 10$^{-3}$, the uncorrected SS-FNS methods show errors below 0.5 eV. The corrected SS-FNS variant further improves the results, and the results converge beyond 10$^{-4}$. These results demonstrate that the SS-FNS framework is reliable for truncating the virtual space in EA calculations in the relativistic framework. For further calculations in this section, we employed a truncation threshold of 10$^{-4.5}$.
\begin{figure}[h]
    \centering
    \includegraphics[width=0.9\textwidth]{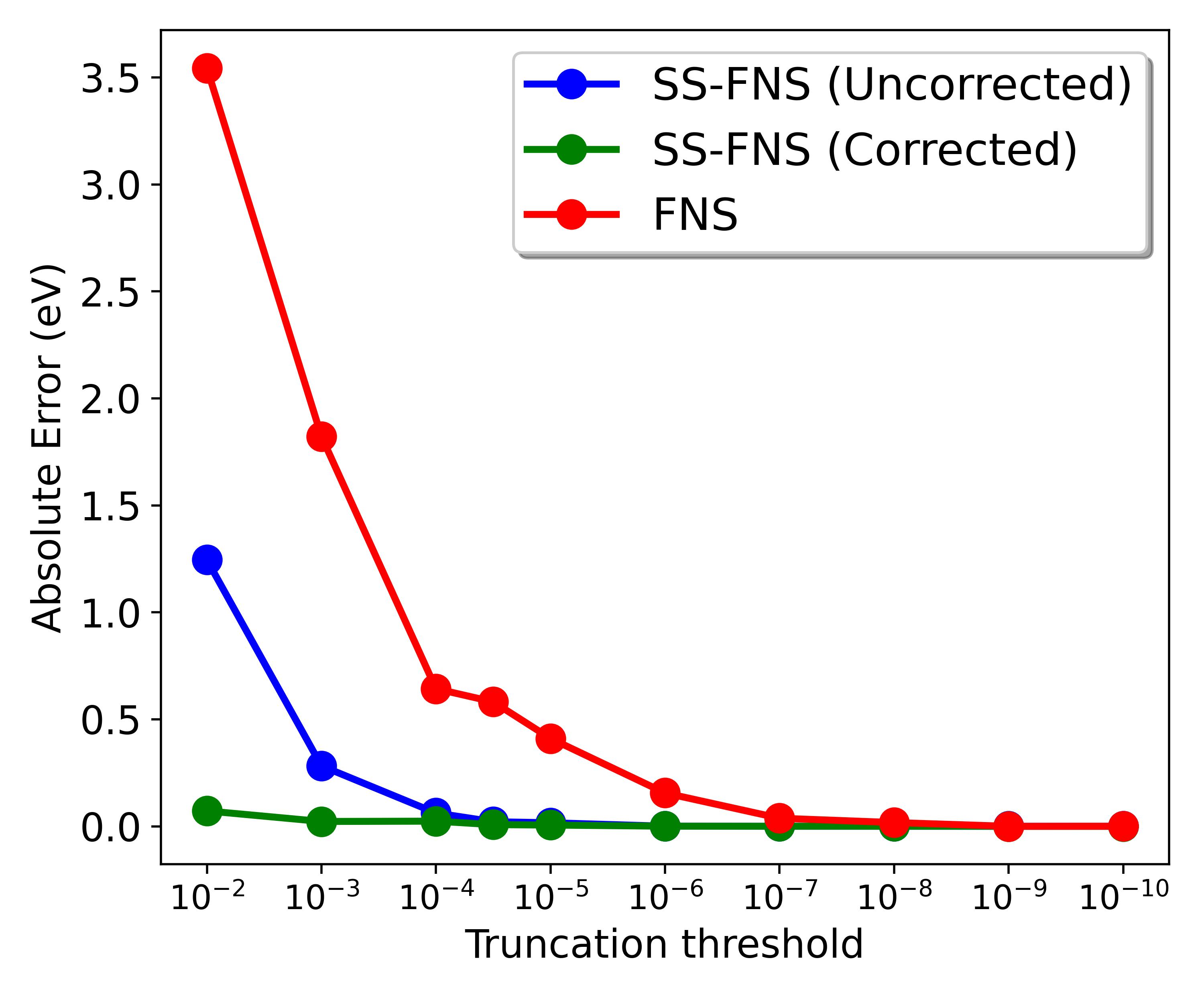} % Replace with your actual path
    % Main caption for the figure
    \caption{\label{fig:ea_thresh_conv}The comparison of absolute error in EA (in eV) for SS-FNS and FNS truncation schemes of CD-based X2CAMF version of EA-ADC(3) with respect to the canonical result for IBr molecule using s-aug-dyall.v2z at different truncation thresholds.}
    \label{fig:ea_thresh_conv}
\end{figure}

\subsubsection{Coin-age metal atoms}
To assess the performance of relativistic ADC(3) employing the X2CAMF Hamiltonian for the EA problem, we have tested the SS-FNS-CD-EA-ADC(3) and the corresponding semi-empirically scaled variant SS-FNS-CD-EA-ADC[(2)+x(3)] for both atomic and molecular systems. For the atomic systems, we have considered coinage metals Cu, Ag, and Au. Table~\ref{table:ea_cuagau} presents the calculated EAs of the coinage metals Cu, Ag, and Au using different variants of the relativistic EA-ADC methods. Specifically, results from the SS-FNS-CD-EA-ADC(3) and SS-FNS-CD-EA-ADC[(2)+x(3)] ($x$=0.5) approaches are compared with canonical EA-ADC(3) and available experimental values. All calculations employed the d-aug-dyall.v4z basis set and the SOC-X2CAMF Hamiltonian, with the SS-FNS threshold set to $10^{-4.5}$. The canonical EA-ADC(3) method gives an EA value of 1.45 eV for Cu. For the SS-FNS-CD-EA-ADC(3), the uncorrected and corrected EA of Cu are 1.39 and 1.45 eV, respectively, indicating a modest but systematic improvement upon applying the perturbative correction. 
\begin{table}[h!]
\centering
\caption{\label{table:ea_cuagau} EA (in eV) of coinage metals using CD-SOCX2CAMF-SS-FNS variant of EA-ADC(3) and EA-ADC[(2)+x(3)] with d-aug-dyall.v4z and x=$0.5$. The canonical EA-ADC(3) calculations are also performed employing CD and X2CAMF Hamiltonian with d-aug-dyall.v4z basis set.}
%\begin{adjustbox}{width=\textwidth}
\vspace{0.5em}
\begin{tabular}{c c c c c c c c c}
\hline
\hline
System&&\multicolumn{2}{c}{\underline{SS-FNS-EA-ADC(3)$^a$}}& \multicolumn{2}{c}{\underline{SS-FNS-EA-ADC[(2)+x(3)]$^a$}}&{\underline{EA-ADC(3)}}&Expt\\
%\cline{3-5} \cline{7-7} 
 &  & Uncorrected& Corrected&Uncorrected &Corrected &Canonical & \\
\hline

Cu && 1.39 & 1.45 &1.13&1.20  & 1.45 & 1.23\cite{bilodeau1998infrared}\\
Ag && 1.21 & 1.26 &1.25&1.30  & 1.27 & 1.30\cite{bilodeau1998infrared}\\
Au && 2.08 & 2.13 &2.27&2.32  & 2.15 & 2.31\cite{gantefor1992zero}\\
\hline
\hline

\end{tabular}
\begin{flushleft}
$^a$ With $10^{-4.5}$ FNS threshold.
\end{flushleft}
%\end{adjustbox}
\end{table}
However, the SS-FNS-CD-EA-ADC(3) method shows an error of 0.21 eV with respect to the experiment for Cu. For Ag, the SS-FNS-CD-EA-ADC(3) method shows improved performance, and the error with respect to the experiment is 0.04 eV. In contrast, the deviation from the canonical EA-ADC(3) result is minimal, only about 0.01 eV. In the case of Au, both the uncorrected and corrected values slightly underestimate the experimental reference and show errors of 0.23 eV and 0.18 eV, respectively.  In all cases, the computed values show excellent agreement with the corresponding canonical ADC(3) results. Notably, the SS-FNS-EA-ADC[(2)+$x$(3)] approach also reproduces the experimental data with high accuracy, exhibiting a maximum deviation of only 0.03 eV.

\begin{table}[h!]
\centering
\caption{\label{table:ea_agx} EA (in eV) of silver monohalides AgX ( X = Cl, Br, I) and AuCl using SS-FNS variant of EA-ADC(3) and EA-ADC[(2)+x(3)] with s-aug-dyall.v4z and x=$0.5$. Experimental values are obtained from Ref.\cite{wu2011photoelectron}. The SS-FNS calculations are performed employing CD and X2CAMF Hamiltonian. }
%\begin{adjustbox}{width=\textwidth}
\vspace{0.5em}
\begin{tabular}{c c c c c c c  c}
\hline
\hline
System&&\multicolumn{2}{c}{\underline{SS-FNS-EA-ADC(3)$^a$}}& \multicolumn{2}{c}{\underline{SS-FNS-EA-ADC[(2)+x(3)]$^a$}}&Expt\\
%\cline{3-5} \cline{7-7} 
 &  & Uncorrected& Corrected&Uncorrected &Corrected & \\
\hline
\hline
AgCl && 1.40 & 1.45 & 1.46 & 1.51    & 1.59\\
AgBr && 1.42 & 1.48 & 1.50  & 1.56    & 1.62\\
AgI  && 1.44 & 1.48 & 1.53  & 1.58   & 1.60\\
AuCl && 1.94 & 1.98 & 2.05  & 2.09    & 2.22\\
\hline\hline

\end{tabular}
\begin{flushleft}
\hspace{1.6cm}$^a$ With $10^{-4.5}$ FNS threshold.
\end{flushleft}
%\end{adjustbox}
\end{table}

\subsubsection{Silver monohalides and AuCl}
For the molecular systems, we have chosen silver monohalides AgX (X = Cl, Br and I) and AuCl. Table~\ref{table:ea_agx} reports the calculated electron affinities (EAs) of the silver halides (AgCl, AgBr, AgI) and gold chloride (AuCl) obtained using the CD-X2CAMF-SS-FNS variants of EA-ADC(3) and EA-ADC[(2)+x(3)] ($x=0.5$) with the s-aug-dyall.v4z basis set with an SS-FNS threshold of $10^{-4.5}$. Experimental reference values were taken from the photoelectron spectroscopy measurements of Wu et al.\cite{wu2011photoelectron}. For the AgX series, both SS-FNS-CD-EA-ADC(3) and SS-FNS-CD-EA-ADC[(2)+x(3)] methods give good agreement with experimental electron affinities with excellent accuracy, although systematic differences between the two methods are evident. The corrected SS-FNS-CD-EA-ADC(3) values are consistently lower than the corresponding SS-FNS-CD-EA-ADC[(2)+x(3)] results by approximately 0.05–0.07 eV across the series. This upward shift introduced by the scaled third-order contribution in EA-ADC[(2)+x(3)] brings the results into closer agreement with experiment. For the three AgX systems, SS-FNS-CD-EA-ADC(3) shows a consistent underestimation of approximately 0.14 eV, which is reduced to less than 0.1 eV upon moving to the ADC[(2)+x(3)] variant. For AuCl, the difference between the two approaches follows a similar trend. The corrected SS-FNS-CD-EA-ADC(3) value of 1.98 eV slightly underestimates the experimental electron affinity (2.22 eV), whereas the EA-ADC[(2)+x(3)] result (2.09 eV) narrows this deviation to less than 0.15 eV. The larger electron affinity of AuCl relative to AgCl primarily arises from the stronger relativistic stabilization of the $6s$ orbital in Au, which enhances the electron binding energy. It is noteworthy that for both atomic and molecular systems, the SS-FNS-CD-EA-ADC[(2)+x(3)] methodology yields quantitatively more accurate electron affinities of heavy-element-containing systems compared with the standard EA-ADC(3) approach.

\subsection{Computational Efficiency }

To evaluate the computational efficiency of the developed relativistic ADC(3) framework, wall-clock times were recorded for IP, EA, and EE calculations of the IBr molecule. All computations were performed on a single node equipped with an Intel(R) Xeon(R) Gold 5315Y CPU @3.20 GHz and 512 GB of RAM, running under identical numerical and convergence settings. The X2CAMF Hamiltonian and the s-aug-dyall.v2z basis set were employed consistently. Canonical as well as FNS (for IP) and SS-FNS (for EA and EE) formulations were executed within the CD–X2CAMF–ADC(3) implementation. As shown in Figure \ref{fig:all_times}, the FNS/SS-FNS approaches offer a remarkable reduction in computational cost compared with their canonical counterparts. For the IP-ADC(3) calculation, the FNS formulation achieves nearly a sixfold speedup, reducing the total wall time from approximately 23 minutes to under 4 minutes. The improvement becomes even more pronounced for the EA-ADC(3) and EE-ADC(3) methods, where the SS-FNS variant shortens the computational time by factors of approximately 15 and 10, respectively. This efficiency gain originates primarily from the subspace-restricted integral transformation and reduced tensor contractions inherent to the FNS/SS-FNS schemes, which avoid the explicit construction of large canonical intermediates while maintaining accuracy. Overall, these results demonstrate that the FNS and SS-FNS formulations significantly enhance the scalability and practical feasibility of high-level relativistic ADC(3) methods for large-scale electron detachment, attachment, and excitation problems, without compromising the accuracy inherent in the theoretically rigorous canonical formalism for dealing systems with heavy elements.

\begin{figure}[h]
    \centering
    \includegraphics[width=1.0\textwidth]{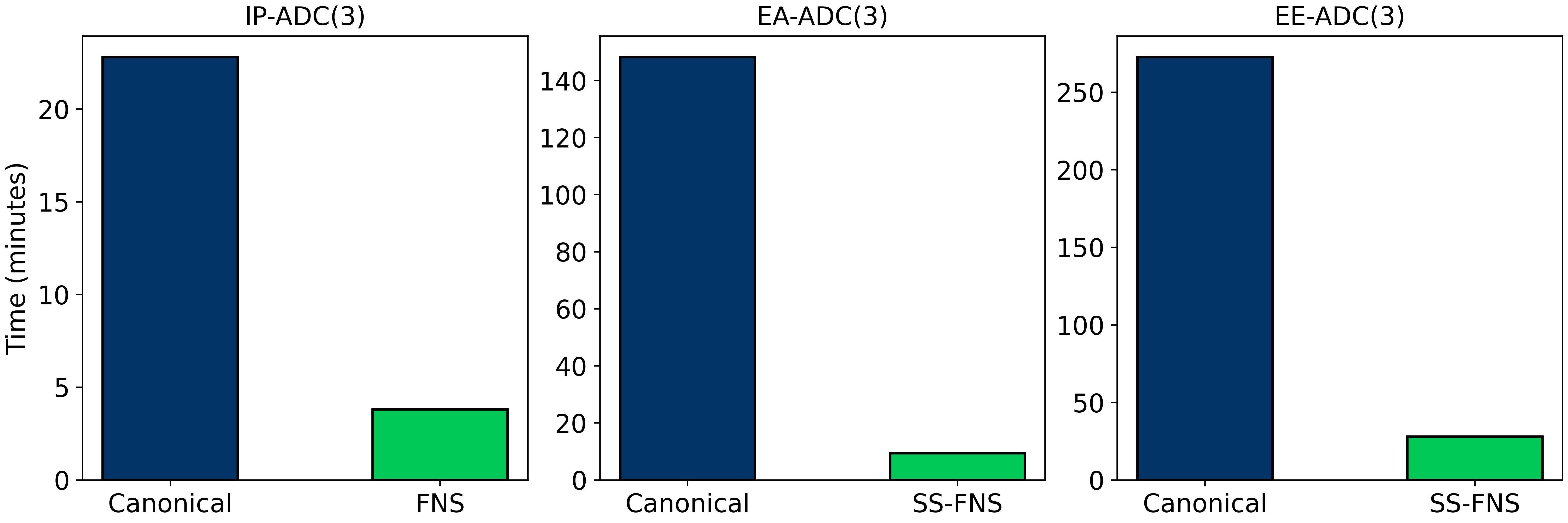} % Replace with your actual path
    % Main caption for the figure
    \caption{\label{fig:all_times}Total wall time taken for IP, EA, and EE-ADC(3) method for IBr molecule in canonical and FNS (for IP) or SS-FNS (for EA and EE) basis using s-aug-dyall.v2z basis set. Both canonical and FNS/SS-FNS calculations are performed using the X2CAMF Hamiltonian and employing CD.}
    \label{fig:all_times}
\end{figure}

\subsection{Application to medium and large systems}

\begin{figure}[h!]
    \centering
    % First subfigure
    \begin{subfigure}[b]{0.5\textwidth}
        \centering
        \includegraphics[width=\textwidth]{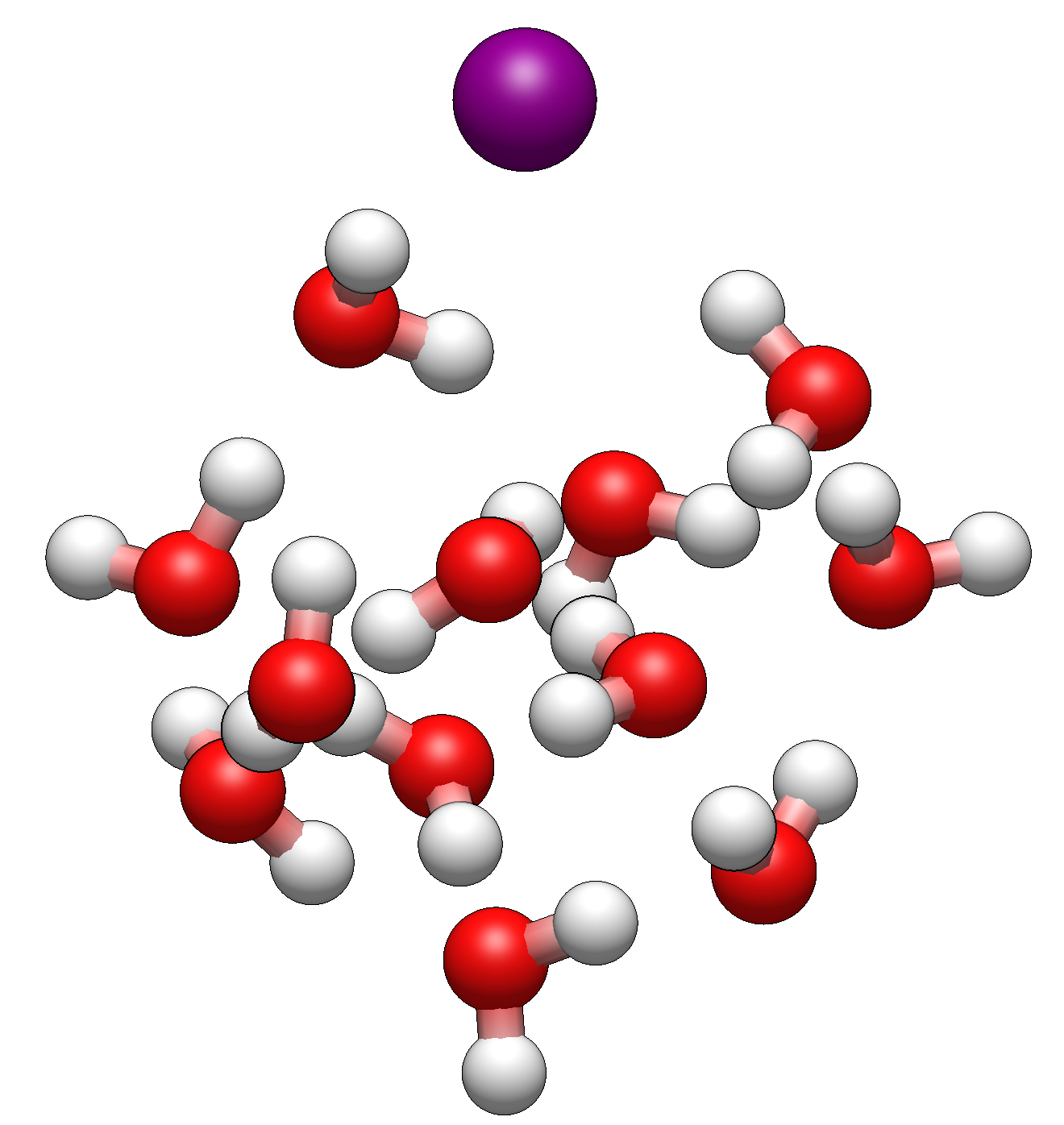}
        \caption{}
        \label{fig:a}
    \end{subfigure}
    \hfill
    % Second subfigure
    \begin{subfigure}[b]{0.45\textwidth}
        \centering
        \includegraphics[width=\textwidth]{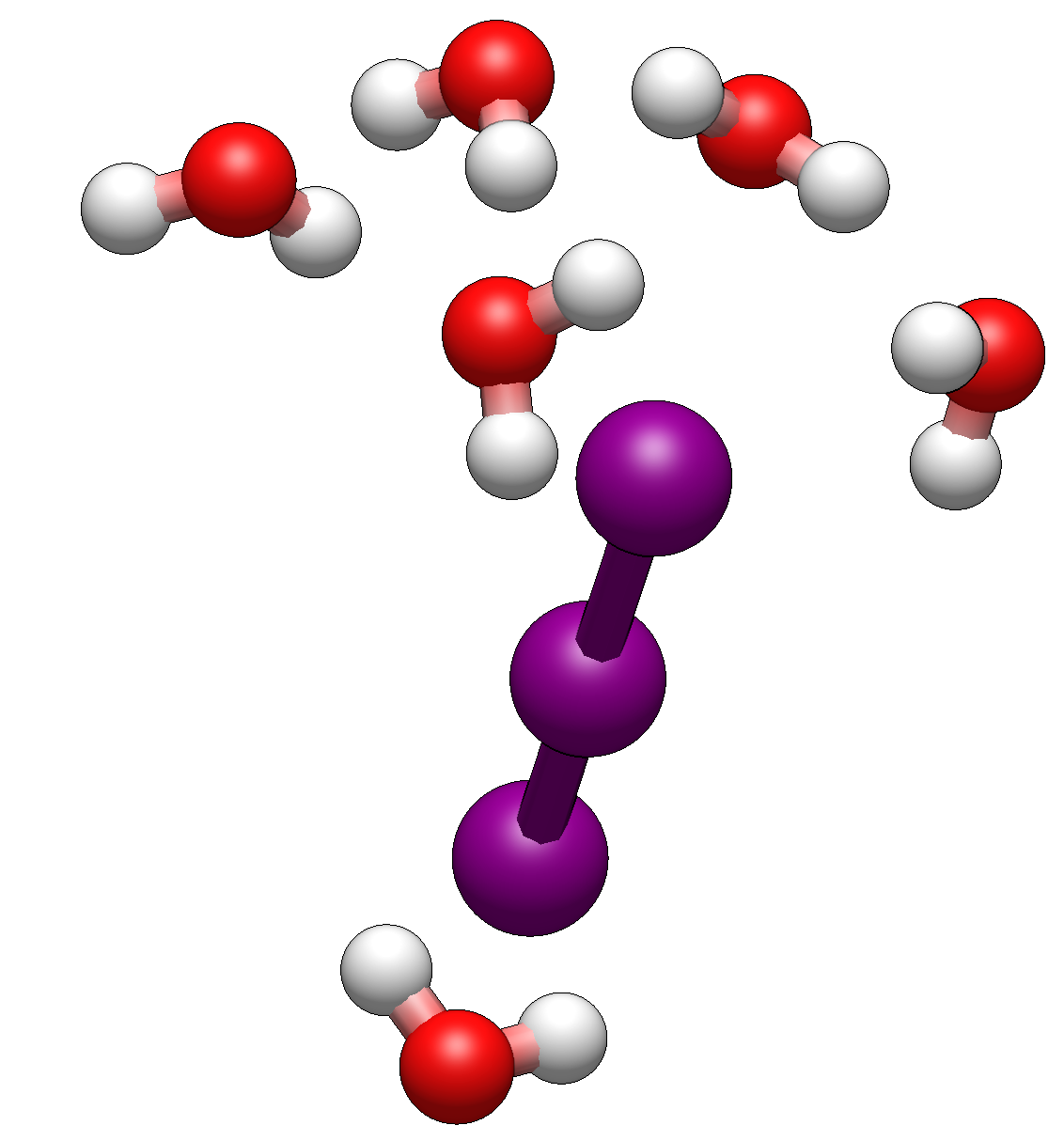}
        \caption{}
        \label{fig:b}
    \end{subfigure}

    \begin{subfigure}[b]{0.66\textwidth}
        \centering
        \includegraphics[width=\textwidth]{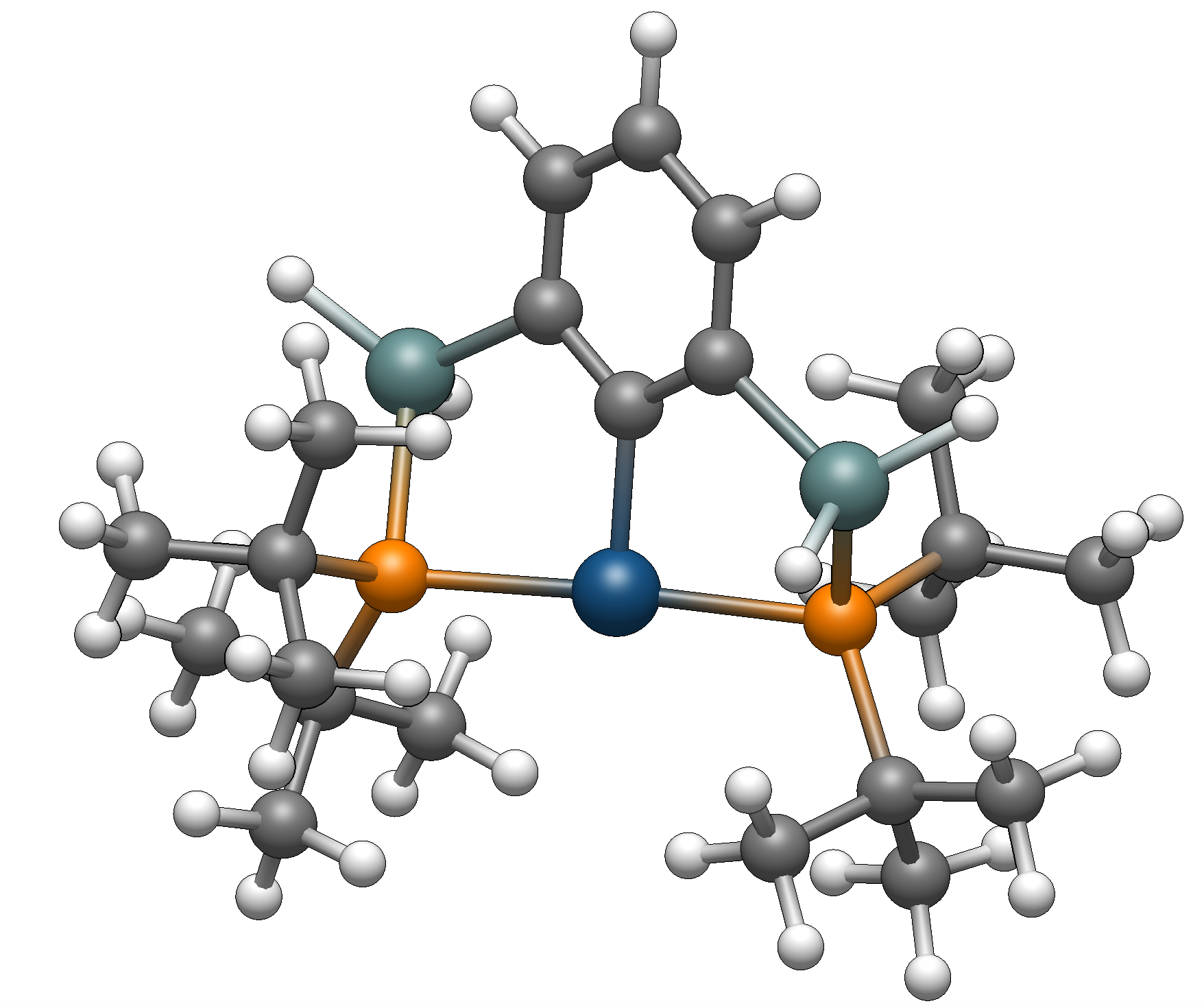}
        \caption{}
        \label{fig:c}
    \end{subfigure}
    
    \caption{(a) \ce{[I(H_{2}O)_{12}]^-} complex, (b) \ce{[I_{3}(H_{2}O)_{6}]^-} complex, (c) Three-coordinated Pincer-Ligated \ce{d^8-Ir(^{tBu4}PGeCGeP)} complex where C is a 2,6-disubstituted phenyl ring}
    \label{fig:1x-complex}
\end{figure}

To assess the efficiency of our newly developed approaches, FNS-CD-IP-ADC(3) and SS-FNS-CD-EE-ADC(3), we performed benchmark studies on medium- and large-sized systems. For the ionization potential (IP) case, two representative examples were chosen: (i) the hydrated anion \ce{[I(H_{2}O)_{12}]^-}, and (ii) a three-coordinate pincer-ligated \ce{d^8-Ir(^{tBu4}PGeCGeP)} complex, where the central carbon belongs to a 2,6-disubstituted phenyl ring (Figure~\ref{fig:1x-complex}).

The structure of the \ce{[I(H_{2}O)_{12}]^-} complex was optimized using DFT at the B3LYP level, including scalar relativistic effects via the zero-order regular approximation (ZORA), as implemented in ORCA 6.0. The SARC-ZORA-TZVP basis set was employed for iodine, while the def2-TZVP basis was used for hydrogen and oxygen. For subsequent FNS-CD-IP-ADC(3) calculations, we used an uncontracted aug-cc-pVDZ basis set for H and O atoms and the s-aug-dyall.v4z basis for I. This results in a total of 1698 virtual spinors. The use of the FNS approximation leads to 114 occupied and 648 virtual spinors, and an auxiliary basis set dimension of 2,574. All computations were carried out sequentially on a workstation with dual Intel(R) Xeon(R) Gold 5315Y processors (3.20 GHz) and 2.0 TB of RAM. The formation of Cholesky vectors in the AO basis took 23 minutes 29 seconds, and the construction of FNS-based two-electron integrals required 1 hour 16 minutes 33 seconds. The ADC(3) step for computing the four lowest vertical IPs consumed 1 day 9 hours 45 minutes 50 seconds, yielding a total wall time of 1 day 16 hours 33 minutes 53 seconds. For comparison, the corresponding FNS-CD-IP-EOM-CCSD calculation required a significantly longer wall time of 4 days 19 hours 5 minutes.\cite{chamoli2025reduced} The first vertical IP obtained from FNS-CD-IP-ADC(3) was 4.22 eV, while FNS-CD-IP-EOM-CCSD gave 4.30 eV. This demonstrates that the ADC(3)-based framework achieves a nearly threefold reduction in computational cost compared with to an EOM-CCSD-based framework.

To further test a larger system, we have chosen a three-coordinate pincer-ligated \ce{d^8}-Ir($^{tBu_{4}}$PGeCGeP) complex, where the central carbon belongs to a 2,6-disubstituted phenyl ring and it has relevance to many areas of modern transition metal chemistry.\cite{van2013pincer, van2012organometallic} As tridentate chelates, pincer ligands offer several channels to tune the electronic and steric environments of unsaturated metal fragments.\cite{roddick2012tuning} In the present work, we have used FNS-CD-IP-ADC(3) to calculate IP of the \ce{Ir(^{R4}PXCXP)} complex, where R is a t-butyl group and X corresponds to a \ce{GeH_{2}} fragment, which acts as a linker. The \ce{Ir(^{R4}PXCXP)} complex possesses a $C_{2}$ equilibrium geometry and the coordinates used for the present FNS-CD-IP-ADC(3) calculation have been taken from Ref. \citenum{baroudi2014calculation}. We have used the dyall.v2z basis set for Ir, Ge, and P, and uncontracted cc-pVDZ basis set for C and H, resulting in a total of 2650 basis functions with 346 occupied and 2304 virtual spinors. For the FNS-CD-IP-ADC(3) calculation, we froze 180 electrons and correlated 166 electrons. After employing FNS threshold of $10^{-4}$ and a cholesky decomposition threshold of $10^{-3}$, resulting in 4120 Cholesky vectors, 166 occupied and 1018 virtual spinors. With this correlation space, the time taken to solve only the ADC(3) secular equation using Davidson is 4 days, 54 minutes, and 56 seconds to calculate one IP state. The value obtained for the first IP of the \ce{Ir(^{R4}PXCXP)} complex is 6.76 eV, resulting from the removal of an electron from the valence $d_{z^2}$-type MO with  $^2A$ symmetry.

To showcase the performance of our SS-FNS-EE-ADC(3) implementation, we have considered the \ce{[I_{3}(H_{2}O)_{6}]^-} complex, and the excitation energy corresponding to the lowest excited state has been calculated. The geometry of the hexa-aqua triiodide complex was obtained from Ref.\citenum{mukhopadhyay2025reduced}, where the same system was used to demonstrate the performance of the SS-FNS-EE-EOM-CCSD method. The dyall.av3z basis set was used for the I atoms, and the uncontracted version of the cc-pVDZ basis set was used for the O and H atoms. The hexa-aqua triiodide complex consists of 21 atoms with 220 electrons. The aforementioned basis set utilizes 1394 virtual spinors in the canonical basis, and a Cholesky threshold of $10^{-3}$ was employed, resulting in 2580 Cholesky vectors. The SS-FNS-EE-ADC(3) calculation was performed on a dedicated workstation with two Intel(R) Xeon(R) Gold 5315Y CPUs @ 3.20 GHz and 2 TB of total RAM. After applying frozen-core approximation and using an FNS threshold of $10^{-4.5}$, the correlation space in the truncated basis contains 70 occupied and 517 virtual spinors. The total time taken for this calculation was 3 days, 18 hours, 25 minutes, and 14 seconds, whereas with a similar computational setting, the SS-FNS-EE-EOM-CCSD calculation took 5 days, 1 hour, 41 minutes. The total time taken for the ADC(3) part in the truncated basis is 2 days, 8 hours, 40 minutes, compared with 3 days, 16 hours, 32 minutes taken by the EOM-CCSD version. The excitation energy obtained for the \ce{2_{g}} state of this complex is 2.16 eV.

\section{Conclusion}

In this work, we have developed a low-cost relativistic third-order ADC framework using CD, FNS, and the X2CAMF Hamiltonian to compute ionization, electron affinity, and excitation properties of heavy-element systems. The inclusion of SS-FNS ensures accurate treatment of excited states, while the CD approach drastically reduces memory and storage requirements. 
Across a diverse set of benchmarks, including IPs from the SOC-81 dataset, fine-structure splittings in halogen oxides, EEs of group-13 cations, I$_3^{-}$), and Xe, as well as EAs of coinage-metal atoms, AgX series, and AuCl, these methods consistently reproduce the corresponding canonical reference results with high accuracy.
Importantly, the approach captures spin–orbit coupling effects without prohibitive computational cost, enabling reliable calculations for systems containing heavy elements that are difficult to compute with canonical 4c-ADC(3) implementations. Furthermore, owing to the perturbative ground-state treatment and Hermitian formulation of the ADC framework, we can compute EE and transition properties for various systems at significantly lower computational cost than EOM-CCSD. With the present implementation, we have achieved 6-15 times the speedup over the canonical CD-X2CAMF-based implementation. We have also successfully calculated the IP of a three-coordinated pincer-ligated complex comprising more than 2600 basis functions, where even after truncation, the virtual space dimension exceeded 1000. These results clearly demonstrate both the efficiency and robustness of the present relativistic ADC(3) implementation in tackling large-scale systems. The results show that the accuracy of the standard relativistic ADC(3) method is inferior to that of the EOM-CCSD method at the same level of truncation for IP, EA, and EE. The use of the semi-empirically scaled variant can improve the accuracy of ADC(3) methods in certain cases. However, their performance is not always consistent. Therefore, it is necessary to develop more accurate approximations within the ADC framework. Work is in progress in this direction.

\section{Acknowledgment}
\sloppy
AKD, SC, and KM acknowledge the support from IIT Bombay, ANRF(CRG/2023/002558), and ISRO  for financial support. The authors also acknowledge the IIT Bombay supercomputing facility and C-DAC (Param Smriti and Param Bramha) for providing computational resources. SC acknowledges the Prime Minister's Research Fellowship (PMRF), and KM acknowledges CSIR-HRDG for the research fellowship. AKD acknowledges the research fellowship funded by the EU NextGenerationEU through the Recovery and Resilience Plan for Slovakia under project No. 09I03-03-V04-00117.

\bibliography{reference}

\end{document}